\documentclass[12pt,a4paper]{article}
\usepackage[utf8]{inputenc}
\usepackage{amsmath}
\usepackage{amsfonts} 
\usepackage{mathtools} 
\usepackage{amssymb} 
\usepackage{subfigure} 
\usepackage{graphicx}
\usepackage{xcolor}
\usepackage{cite} 
\usepackage{mathrsfs} 
\usepackage{slashbox}
\usepackage{multirow}

\graphicspath{{Figs/}}

\newcommand{\abs}[1]{|#1|} 
\newcommand{\ABS}[1]{\left|#1\right|} 

\newcommand{\refEQ}[1]{eq.\,\eqref{#1}} 
\newcommand{\refEQS}[1]{eqs.\,\eqref{#1}} 

\newcommand{\WL}{W_{\rm L}}\newcommand{\WLd}{\WL^\dagger}
\newcommand{\WqR}[1]{W_{\rm {#1}_R}}\newcommand{\WqRd}[1]{\WqR{#1}^\dagger}
\newcommand{\WuR}{\WqR{u}}\newcommand{\WuRd}{\WqRd{u}}
\newcommand{\WdR}{\WqR{d}}\newcommand{\WdRd}{\WqRd{d}}
\newcommand{\UqX}[2]{\mathcal U_{#2}^{#1}}\newcommand{\UqXd}[2]{\mathcal U_{#2}^{#1\dagger}}
\newcommand{\UuL}{\UqX{u}{L}}\newcommand{\UuLd}{\UqXd{u}{L}}
\newcommand{\UdL}{\UqX{d}{L}}\newcommand{\UdLd}{\UqXd{d}{L}}
\newcommand{\UuR}{\UqX{u}{R}}\newcommand{\UuRd}{\UqXd{u}{R}}
\newcommand{\UdR}{\UqX{d}{R}}\newcommand{\UdRd}{\UqXd{d}{R}}

\newcommand{\Yd}[1]{\Gamma_{#1}}
\newcommand{\Yu}[1]{\Delta_{#1}}
\newcommand{\Ydd}[1]{\Gamma_{#1}^\dagger}
\newcommand{\Yud}[1]{\Delta_{#1}^\dagger}
\newcommand{\PR}[1]{{\rm P}_{\!#1}}

\newcommand{\PRXL}[2]{P_{#1}^{[#2_L]}}
\newcommand{\PRXR}[2]{P_{#1}^{[#2_R]}}

\newcommand{\wLP}[1]{{\rm L_{#1}^0}}
\newcommand{\LP}[1]{{\rm L_{#1}}}
\newcommand{\wRP}[1]{{\rm R_{#1}^0}}
\newcommand{\RP}[1]{{\rm R_{#1}}}
\newcommand{\id}{\mathbf{1}}

\newcommand{\unq}[2]{\hat n_{{\rm [#1]}#2}^{\phantom{\ast}}}
\newcommand{\unqC}[2]{\hat n_{{\rm [#1]}#2}^{\ast}}
\newcommand{\und}[1]{\unq{d}{#1}}
\newcommand{\unu}[1]{\unq{u}{#1}}
\newcommand{\undC}[1]{\unqC{d}{#1}}
\newcommand{\unuC}[1]{\unqC{u}{#1}}
\newcommand{\unqvec}[1]{\hat n_{{\rm [#1]}}^{\phantom{\ast}}}

\newcommand{\undvec}{\unqvec{d}}
\newcommand{\unuvec}{\unqvec{u}}

\newcommand{\rnq}[2]{\hat r_{{\rm [#1]}#2}^{\phantom{\ast}}}
\newcommand{\rnqC}[2]{\hat r_{{\rm [#1]}#2}^{\ast}}
\newcommand{\rnd}[1]{\rnq{d}{#1}}
\newcommand{\rnu}[1]{\rnq{u}{#1}}
\newcommand{\rndC}[1]{\rnqC{d}{#1}}
\newcommand{\rnuC}[1]{\rnqC{u}{#1}}
\newcommand{\rnqvec}[1]{\hat r_{{\rm [#1]}}^{\phantom{\ast}}}

\newcommand{\rndvec}{\rnqvec{d}}
\newcommand{\rnuvec}{\rnqvec{u}}

\newcommand{\Hd}[1]{\Phi_{#1}}

\newcommand{\Hdt}[1]{\tilde\Phi_{#1}}
\newcommand{\nHH}{\mathrm{h}^0}
\newcommand{\nHR}{\mathrm{R}^0}
\newcommand{\nHI}{\mathrm{I}^0}

\newcommand{\cH}{H^\pm}

\newcommand{\cHp}{H^+}

\newcommand{\cb}{c_\beta}
\renewcommand{\sb}{s_\beta}

\newcommand{\tb}{t_\beta}
\newcommand{\tbinv}{\tb^{-1}}
\newcommand{\tti}{\tb+\tbinv}

\newcommand{\CKM}{V}\newcommand{\CKMd}{\CKM^\dagger}
\newcommand{\V}[1]{{\CKM_{#1}^{\phantom{\ast}}}}
\newcommand{\Vc}[1]{{\CKM_{#1}^\ast}}

\newcommand{\wMQ}[1]{M_{#1}^0}
\newcommand{\wNQ}[1]{N_{#1}^0}

\newcommand{\mMU}{M_u} 
\newcommand{\wMU}{M_u^0} 
\newcommand{\mMD}{M_d} 
\newcommand{\wMD}{M_d^0} \newcommand{\wMDd}{M_d^{0\dagger}}
\newcommand{\mNU}{N_u} \newcommand{\wNU}{N_u^{0}}
\newcommand{\mNUd}{N_u^{\dagger}} 
\newcommand{\mND}{N_d} \newcommand{\wND}{N_d^{0}}
\newcommand{\mNDd}{N_d^{\dagger}}

\newcommand{\ZZ}[1]{\mathbb{Z}_{#1}}
\newcounter{notas}
\graphicspath{{Figs/}}

\begin{document}
\begin{titlepage}

\hfill\begin{minipage}[r]{0.3\textwidth}\begin{flushright}  CFTP/18-006\\    IFIC/18-xxx \end{flushright} \end{minipage}

\begin{center}

\vspace{0.50cm}

{\large \bf {Symmetry Constrained Two Higgs Doublet Models}}

\vspace{0.50cm}

Jo\~{a}o M. Alves $^{a,}$\footnote{\texttt{j.magalhaes.alves@tecnico.ulisboa.pt}}
Francisco J. Botella $^{b,}$\footnote{\texttt{Francisco.J.Botella@uv.es}}, 
Gustavo C. Branco  $^{a,}$\footnote{\texttt{gbranco@tecnico.ulisboa.pt}},\\ 
Fernando Cornet-Gomez $^{b,}$\footnote{\texttt{Fernando.Cornet@ific.uv.es}}, 
Miguel Nebot $^{a,}$\footnote{\texttt{miguel.r.nebot.gomez@tecnico.ulisboa.pt}}
and Jo\~{a}o P. Silva $^{a,}$\footnote{\texttt{jpsilva@cftp.ist.utl.pt}}
\end{center}

\vspace{0.50cm}
\begin{flushleft}
\emph{$^a$ Departamento de F\'{\i}sica and Centro de F\'{\i}sica Te\'{o}rica de
Part\'{\i}culas (CFTP),\\
\quad Instituto Superior T\'{e}cnico (IST), U. de Lisboa (UL),\\ 
\quad Av. Rovisco Pais, P-1049-001 Lisboa, Portugal.} \\
\emph{$^b$ Departament de F\'{\i}sica Te\`{o}rica and IFIC,\\
\quad Universitat de Val\`{e}ncia-CSIC,\\
\quad E-46100, Burjassot, Spain.} 
\end{flushleft}

\vspace{0.5cm}

\begin{abstract}
We study Two-Higgs-Doublet Models (2HDM) where Abelian symmetries have been introduced, leading to a drastic reduction in the number of free parameters in the 2HDM. Our analysis is inspired in BGL models, where, as the result of a symmetry of the Lagrangian, there are tree-level scalar mediated Flavour-Changing-Neutral-Currents, with the flavour structure depending only on the CKM matrix.
A systematic analysis is done on the various possible schemes, which are classified in different classes, depending on the way the extra symmetries constrain the matrices of couplings defining the flavour structure of the scalar mediated neutral currents. All the resulting flavour textures of the Yukawa couplings are stable under renormalisation since they result from symmetries imposed at the Lagrangian level. We also present a brief phenomenological analysis of the most salient features of each class of symmetry constrained 2HDM.
\end{abstract}

\end{titlepage}

\newpage

\clearpage
\section{Introduction\label{SEC:INTRO}}
One of the simplest extensions of the Standard Model (SM) consists of the introduction of one or more additional scalar doublets to its spectrum. The first 2 Higgs Doublet Model (2HDM) was proposed by Lee \cite{Lee:1973iz} in order to generate spontaneous CP violation, at a time when only two incomplete generations were known. The general 2HDM \cite{Branco:2011iw,Ivanov:2017dad} has a priori two flavour problems:
\begin{enumerate}
\item[(i)] it has potentially dangerous scalar mediated Flavour Changing Neutral Currents (FCNC) at tree level,
\item[(ii)] it leads to a large increase in the number of flavour parameters in the scalar sector, parametrised by two arbitrary $3\times 3$ complex matrices, which we denote by $\mND$ and $\mNU$.
\end{enumerate}
The first problem was elegantly solved by Glashow and Weinberg \cite{Glashow:1976nt} through the introduction of a $\ZZ{2}$ discrete symmetry. However, this $\ZZ{2}$ symmetry renders it impossible to generate either spontaneous or explicit CP violation in the scalar sector, in the context of 2HDM. Both explicit \cite{Weinberg:1976hu} and spontaneous \cite{Branco:1979pv} CP violation in the scalar sector can be obtained if one introduces a third scalar doublet while maintaining FCNC in the scalar sector. Recently, it was pointed out \cite{Branco:2015bfb} that an intriguing correlation exists between the possibility of a given scalar potential to generate explicit and spontaneous CP violation.
Indeed in most examples studied, if a given scalar potential can generate spontaneous CP violation, it can also have explicit CP violation in the scalar sector.

In a separate development, which addresses simultaneously the above two problems of 2HDM, it was shown \cite{Branco:1996bq} by Branco, Grimus and Lavoura (BGL) that one may have a scenario where there are tree level FCNC, but with $\mND$ and $\mNU$ fixed entirely by the elements of the Cabibbo-Kobayashi-Maskawa (CKM) matrix. In some BGL models, the suppression of FCNC couplings resulting from the smallness of CKM elements, is such that the new neutral scalars need not be too massive in order to conform with experiment. BGL models have been studied in the literature \cite{Botella:2009pq,Botella:2011ne} and their phenomenological consequences have been analysed in the context of the LHC \cite{Botella:2014ska,Bhattacharyya:2014nja,Botella:2015hoa}. A generalisation of BGL models has been recently proposed in the framework of 2HDM \cite{Alves:2017xmk}.

Regarding symmetries, Ferreira and Silva \cite{Ferreira:2010ir} classified all possible implementations of Abelian symmetries in 2HDM with fermions which lead to non-vanishing quark masses and a CKM matrix which is not block diagonal (see also \cite{Serodio:2013gka}).
 
In this paper we study in a systematic way scenarios arising from different implementations of Abelian symmetries in the context of 2HDM  which can lead to a natural reduction in the number of parameters in these models. In the search for these scenarios, we were inspired by BGL and generalised BGL (gBGL) models where the coupling matrices $\mND$, $\mNU$ (see \refEQS{eq:MN:matrices:00:2}--\eqref{eq:Yukawa:01}) can be written in terms of the quark mass matrices and projection operators. Thus we classify the different models according to the structures of $\mND$, $\mNU$. We identify the symmetry leading to each of the models and the corresponding flavour textures of the Yukawa couplings. These textures are stable under renormalisation, since they result from symmetries of the Lagrangian. 

The organisation of the paper is the following.
The notation is set up in section~\ref{SEC:GENERALITIES}.
We then present our main results in sections
\ref{SEC:LEFT} and \ref{SEC:RIGHT},
obeying what we denote the \emph{Left} and \emph{Right conditions}
introduced in eqs.~\eqref{eq:LeftProjection:00}
and \eqref{eq:RightProjection:00}, respectively.
We show that,
besides BGL and gBGL there is a new type of model
obeying \emph{Left conditions} and that there are
six classes of models obeying \emph{Right conditions} which,
as far as we can tell,
are presented in full generality here for the first time.
For definiteness, we concentrate on the quark sector.
Some of the most salient phenomenological implications are presented in section~\ref{SEC:PHENO},
and our conclusions appear in section~\ref{SEC:CONC}.
We defer some technical details to appendix~\ref{SEC:APP:}.
In particular, we present in appendix~\ref{sSEC:APP:Conditions}
conditions for the identification of the
various models which are invariant under basis transformations in
the spaces of left-handed doublets and of up-type and down-type
right-handed singlets.

\section{Generalities and notation\label{SEC:GENERALITIES}}
The Yukawa Lagrangian, with summation over fermion generation indices implied and omitted, reads
\begin{equation}\label{eq:Yukawa:00}
\mathscr L_{\rm Y}=
-\bar Q_L^0[\Yd{1}\Hd{1}+\Yd{2}\Hd{2}] d_R^0 -\bar Q_L^0[\Yu{1}\Hdt{1}+\Yu{2}\Hdt{2}] u_R^0+\text{H.c.},
\end{equation}
with $\Hdt{j}=i\sigma_2\Hd{j}^\ast$. Electroweak spontaneous symmetry breaking arises from the vacuum expectation values of the scalar doublets
\begin{equation}\label{eq:EWSSB:00}
\langle\Hd{1}\rangle=\begin{pmatrix} 0\\ e^{i\xi_1}v_1/\sqrt{2}\end{pmatrix},\quad \langle\Hd{2}\rangle=\begin{pmatrix} 0\\ e^{i\xi_2}v_2/\sqrt{2}\end{pmatrix}.
\end{equation}
We use $v^2\equiv v_1^2+v_2^2$, $\cb=\cos\beta\equiv v_1/v$, $\sb=\sin\beta\equiv v_2/v$, $\tb\equiv\tan\beta$ and $\xi\equiv\xi_2-\xi_1$. In the ``Higgs basis'' \cite{Georgi:1978ri,Donoghue:1978cj,Botella:1994cs}
\begin{equation}\label{eq:HiggsBasis:00}
\begin{pmatrix}H_{1}\\ H_{2}\end{pmatrix}=
\begin{pmatrix}\cb & \phantom{-}\sb\\ \sb & -\cb \end{pmatrix}
\begin{pmatrix}e^{-i\xi_1}\Hd{1}\\ e^{-i\xi_2}\Hd{2}\end{pmatrix},
\end{equation}
only $H_{1}$ has a non-zero vacuum expectation value 
\begin{equation}\label{eq:HiggsBasis:01}
\langle H_{1}\rangle=\begin{pmatrix} 0\\ v/\sqrt{2}\end{pmatrix},\quad \langle H_{2}\rangle=\begin{pmatrix} 0\\ 0\end{pmatrix}.
\end{equation}
Expanding the scalar fields around \refEQ{eq:HiggsBasis:01}, one has
\begin{equation}\label{eq:HiggsBasis:02}
H_{1}=\begin{pmatrix} G^+\\ (v+\nHH+iG^0)/\sqrt{2}\end{pmatrix},\quad H_{2}=\begin{pmatrix} \cHp\\ (\nHR+i\nHI)/\sqrt{2}\end{pmatrix},
\end{equation}
with $G^0$, $G^\pm$ the would-be Goldstone bosons, $\nHH$, $\nHR$, $\nHI$ neutral fields and $\cH$ the charged scalar. 
Then, the Yukawa couplings in \refEQ{eq:Yukawa:00} read
\begin{equation}
-\frac{v}{\sqrt 2}\mathscr L_{\rm Y}=
\bar Q_L^0(\wMD H_{1}+\wND H_{2}) d_R^0 +\bar Q_L^0(\wMU \tilde H_{1}+\wNU \tilde H_{2}) u_R^0+\text{H.c.},
\end{equation}
with the $\wMD$, $\wMU$ mass matrices, and the $\wND$, $\wNU$ matrices given by
\begin{alignat}{3}
\wMD&=\frac{ve^{i\xi_1}}{\sqrt 2}(\cb\Yd{1}+e^{i\xi}\sb\Yd{2})\, ,&\quad \wND&=\frac{ve^{i\xi_1}}{\sqrt 2}(\sb\Yd{1}-e^{i\xi}\cb\Yd{2})\, ,
\label{eq:MN:matrices:00:1}\\
\wMU&=\frac{ve^{-i\xi_1}}{\sqrt 2}(\cb\Yu{1}+e^{-i\xi}\sb\Yu{2})\, ,&\quad \wNU&=\frac{ve^{-i\xi_1}}{\sqrt 2}(\sb\Yu{1}-e^{-i\xi}\cb\Yu{2})\,.
\label{eq:MN:matrices:00:2}
\end{alignat}
This Lagrangian can be written in terms of physical quantities as
\begin{equation}\label{eq:Yukawa:01}
-\frac{v}{\sqrt 2}\mathscr L_{\rm Y}
=(\bar u_L\CKM,\bar d_L)(\mMD H_{1}+\mND H_{2}) d_R +(\bar u_L,\bar d_L\CKMd)(\mMU \tilde H_{1}+\mNU \tilde H_{2}) u_R+\text{H.c.}
\end{equation}
We have used the
usual bidiagonalisations into the mass bases,
\begin{equation}\label{eq:Diagonalisation}
\UdLd\wMD\,\UdR=\mMD=\text{diag}(m_d,m_s,m_b),\quad\UuLd\wMU\,\UuR=\mMU=\text{diag}(m_u,m_c,m_t),
\end{equation}
via $3\times 3$ unitary matrices $\UqX{q}{X}$ ($q=u,d$, $X=L,R$).
$\CKM=\UuLd\UdL$ in \refEQ{eq:Yukawa:01} is the CKM mixing matrix.
While the quark masses $\mMD$ and $\mMU$ in \refEQ{eq:Diagonalisation} are characterised by $3+3=6$ physical parameters, in a general 2HDM the complex matrices
\begin{equation}\label{eq:Diagonalisation:Nq}
\UdLd\wND\,\UdR=\mND,\quad\UuLd\wNU\,\UuR=\mNU,
\end{equation}
are free. This introduces in principle $2\times 3\times 3\times 2=36$ new real parameters\footnote{Notice however that the bidiagonalisation of the mass matrices still leaves the freedom to rephase individual quark fields. Together with the CKM matrix, the $\mND$, $\mNU$ matrices should enter physical observables in rephasing invariant combinations
\cite{Botella:1994cs}.}. This large freedom is certainly a source of concern since, for example, FCNC can put significant constraints on $\mND$ and $\mNU$.\\
Invariance under some (symmetry) transformation is the best motivated requirement which can limit this inflation of parameters.
Following \cite{Ferreira:2010ir},
we consider in particular Abelian symmetry transformations
\begin{equation}\label{eq:Charges:00}
\Hd{1}\mapsto\Hd{1},\ \Hd{2}\mapsto e^{i\theta}\Hd{2},\ Q_{Lj}^0\mapsto e^{i\alpha_j\theta}Q_{Lj}^0,\ d_{Rj}^0\mapsto e^{i\beta_j\theta}d_{Rj}^0,\ u_{Rj}^0\mapsto e^{i\gamma_j\theta}u_{Rj}^0\,,
\end{equation}
where $\alpha_j$, $\beta_j$, $\gamma_j$, are the charges of the different fermion doublets and singlets normalized to the charge of the second scalar doublet $\Hd{2}$. As already mentioned, all possible realistic implementations of \refEQ{eq:Charges:00} were classified in \cite{Ferreira:2010ir}. In BGL models and their generalization in \cite{Alves:2017xmk}, the symmetry properties had an interesting translation into relations among the $\wNQ{q}$ and $\wMQ{q}$ matrices (very useful for example in the study of the renormalization group evolution of the Yukawa matrices). Having such a connection between a symmetry and matrix relations is not always possible. Inspired by the existence of that property in those two interesting classes of models, we focus on 2HDMs which obey an Abelian symmetry, \refEQ{eq:Charges:00}, and which fulfill an additional requirement; either (a) or (b) below:
\begin{itemize}
\item[(a)] The Yukawa coupling matrices are required to obey \emph{Left conditions} 
\begin{equation}\label{eq:LeftProjection:00}
\wND=\wLP{d}\,\wMD\,,\quad \wNU=\wLP{u}\,\wMU\,,
\end{equation}
with
\begin{equation}\label{eq:LP:00}
\wLP{q}=\ell_1^{[q]}\PR{1}+\ell_2^{[q]}\PR{2}+\ell_3^{[q]}\PR{3}\,,
\end{equation}
where $\ell_j^{[q]}$ are, \emph{a priori}, arbitrary numbers.
Here and henceforth we shall often use the index $q$ to refer
to matrices in the up ($q=u$) or down ($q=d$) sectors.
We have used the projection operators $\PR{i}$ 
defined by $[\PR{i}]_{jk}=\delta_{ij}\delta_{jk}$ (no sum in $j$).
In matrix form:
\begin{equation}
P_1 =
\left(
\begin{array}{ccc}
1 & 0 & 0\\
0 & 0 & 0\\
0 & 0 & 0
\end{array}
\right),
\ \ 
P_2 =
\left(
\begin{array}{ccc}
0 & 0 & 0\\
0 & 1 & 0\\
0 & 0 & 0
\end{array}
\right),
\ \ 
P_3 =
\left(
\begin{array}{ccc}
0 & 0 & 0\\
0 & 0 & 0\\
0 & 0 & 1
\end{array}
\right).
\label{P_i:MatrixForm}
\end{equation}
These projection operators satisfy
$\PR{i}\PR{j}=\delta_{ij}\PR{i}$ (no sum in $i$) and $\sum_i\PR{i}=\id$. 
\item[(b)] The Yukawa coupling matrices are instead required to obey \emph{Right conditions}
\begin{equation}\label{eq:RightProjection:00}
\wND=\wMD\,\wRP{d}\,,\quad \wNU=\wMU\,\wRP{u}\,,
\end{equation}
with
\begin{equation}\label{eq:RP:00}
\wRP{q}=r_1^{[q]}\PR{1}+r_2^{[q]}\PR{2}+r_3^{[q]}\PR{3}\,,
\end{equation}
where $r_j^{[q]}$ are, again \emph{a priori}, arbitrary numbers and, as in \refEQ{eq:LP:00},
$\PR{i}$ are the projection operators in eqs.~\eqref{P_i:MatrixForm}.
\end{itemize}
Upper (and lower) case L's and R's are used in correspondence with the matrices
(and parameters) acting on the left or the right of $\wMQ{q}$ in \refEQS{eq:LeftProjection:00} and \eqref{eq:RightProjection:00}. Although it is not required \emph{a priori}, the matrices $\wLP{q}$ and $\wRP{q}$ are non-singular.\\
All the resulting models, that is all 2HDMs obeying \refEQ{eq:Charges:00} and either \emph{Left} or \emph{Right conditions} are analysed in section~\ref{SEC:LEFT} and section~\ref{SEC:RIGHT}, respectively.\\ 
We emphasize that our aim is to reduce the number of parameters.
As shown in ref.~\cite{Ferreira:2010ir},
imposing Abelian symmetries leaves only a reduced set of possible models,
each with a significantly reduced number of independent parameters.
Here, we consider only those Abelian models which
can in some sense be seen as generalizations of the BGL models,
by imposing, in addition,
the \emph{Left conditions} in eq.~\eqref{eq:LeftProjection:00},
or the \emph{Right conditions} in eq.~\eqref{eq:RightProjection:00}.
As anticipated,
the number of independent parameters of the models is significantly
reduced with respect to the most general 2HDM. 
It is to be noticed that,
as shown in sections \ref{sSEC:LEFT:ell_i} and \ref{sSEC:RIGHT:r_i},
$\ell_j^{[q]}$ or $r_j^{[q]}$, which are a priori arbitrary,
turn out to be unavoidably fixed in terms of $\tb$.
Quite significantly, as analysed in appendix \ref{sSEC:APP:RowsColumns},
\refEQS{eq:LeftProjection:00} and \eqref{eq:RightProjection:00}
have an elegant interpretation.
In the popular 2HDMs of types I, II, X and Y  \cite{Haber:1978jt,Donoghue:1978cj,Hall:1981bc,Barger:1989fj},
a $\ZZ{2}$ symmetry is incorporated and it eliminates the possibility
of FCNC.
But, in those cases,
the $\ZZ{2}$ assignment is universal for the different fermion
families;
all fermions of a given charge couple to the \emph{same} scalar doublet.
Here, \refEQS{eq:LeftProjection:00} and \eqref{eq:RightProjection:00} have a different non-universal interpretation which leads to controlled FCNC:
\begin{itemize}
\item in the models of section \ref{SEC:LEFT}, obtained by imposing the \emph{Left conditions} in \refEQ{eq:LeftProjection:00}, each left-handed doublet $Q_{Li}^0$ couples exclusively, i.e. to one and only one, of the scalar doublets $\Hd{k}$,
\item in the models of section \ref{SEC:RIGHT}, obtained by imposing the \emph{Right conditions} in \refEQ{eq:RightProjection:00}, each right-handed singlet $d_{Ri}^0$, $u_{Rj}^0$, couples exclusively to one scalar doublet $\Hd{k}$.
\end{itemize}
In particular, we stress that here, and in contrast to type I, II, X, and Y models,
fermions of a given electric charge but different families need not couple
all to the same scalar doublet.
In this sense,
conditions \eqref{eq:LeftProjection:00} and \eqref{eq:RightProjection:00}
- applied in the context of models with Abelian symmetries - can also
be seen as a generalization of the Glashow,
Weinberg conditions \cite{Glashow:1976nt}
for Natural Flavour Conservation (NFC).
In the present approach,
having $L^0_d$ and $L^0_u$ proportional to the identity
(or $R^0_d$ and $R^0_u$ proportional to the identity)
enforces the NFC type I and type II 2HDM.

\section{Symmetry Controlled Models with ``Left'' Conditions\label{SEC:LEFT}}
We present in this section the different models arising from an Abelian symmetry and for which there are matrices $\wLP{d}$ and $\wLP{u}$ such that \refEQ{eq:LeftProjection:00} is verified. 
To this end, we have constructed a program which produces \emph{all}
models satisfying the Abelian symmetries in \refEQ{eq:Charges:00},
and which lead to non-vanishing quark masses and a
CKM matrix which is not block diagonal,
thus verifying the results in ref.~\cite{Ferreira:2010ir}.\footnote{In
fact, there is a misprint in eq.~(89) of the published version
of \cite{Ferreira:2010ir}, which however is correct in the arxiv version.}
For each Abelian model, the program then checks if it satisfies in addition
\refEQ{eq:LeftProjection:00}.
Thus, our final list will be complete.
Before addressing the models themselves, it is convenient to make some observations on the effect of rotating into mass bases of the up and down quarks.
\subsection{Conditions in the mass basis\label{sSEC:LEFT:Conditions}}
In the mass bases, given by the unitary transformations in \refEQ{eq:Diagonalisation}, \refEQ{eq:LeftProjection:00} reads
\begin{equation}\label{eq:LeftProjection:01}
\mND=\LP{d}\,\mMD\,,\quad \mNU=\LP{u}\,\mMU\,,
\end{equation}
with the transformed matrices
\begin{equation}
\LP{d}=\UdLd\,\wLP{d}\,\UdL,\quad \LP{u}=\UuLd\,\wLP{u}\,\UuL\,.
\end{equation}
Introducing transformed projection operators
\begin{equation}\label{eq:LP:02}
\PRXL{j}{d}\equiv \UdLd\,\PR{j}\,\UdL,\quad\PRXL{j}{u}\equiv \UuLd\,\PR{j}\,\UuL\,,
\end{equation}
one simply has
\begin{equation}\label{eq:LP:01}
\LP{d}=\ell_1^{[d]}\PRXL{1}{d}+\ell_2^{[d]}\PRXL{2}{d}+\ell_3^{[d]}\PRXL{3}{d}\,,\quad
\LP{u}=\ell_1^{[u]}\PRXL{1}{u}+\ell_2^{[u]}\PRXL{2}{u}+\ell_3^{[u]}\PRXL{3}{u}\,.
\end{equation}
Furthermore, since the CKM matrix is $\CKM=\UuLd\UdL$, one has the straightforward relation
\begin{equation}\label{eq:LP:03}
\PRXL{k}{u}=\CKM\,\PRXL{k}{d}\,\CKMd\,,
\end{equation}
which is relevant for the parametrisation of the FCNC couplings in the discussion to follow.

\subsection{How to determine $\ell_i$\label{sSEC:LEFT:ell_i}}
Here we show how one determines the coefficients $\ell_i$ ($i=1,2,3$)
just by examining the form of the Yukawa matrices $\Yd{1}$ and $\Yd{2}$.
For definiteness, we concentrate on the down sector.
The reasoning for the up sector follows similar lines and yields
the same conclusions.

As a first step, we notice that, under the assumption of an
Abelian symmetry \cite{Ferreira:2010ir},
\begin{equation}
\left( \Yd{1} \right)_{i a} \neq 0
\ \ \Rightarrow \ \ 
\left( \Yd{2} \right)_{i a} = 0,
\end{equation}
(and the converse $1 \leftrightarrow 2$ also holds);
notice that this implication involves the \emph{same} matrix element of $\Yd{1}$ and $\Yd{2}$.

As a second step, consider $\left( \Yd{1} \right)_{i a} \neq 0$.
We already know that this implies $\left( \Yd{2} \right)_{i a} = 0$.
But then,
the $(ia)$ entries in eqs.~\eqref{eq:MN:matrices:00:1} yield
\begin{equation}
\left( \wMD \right)_{ia} = \frac{ve^{i\xi_1}}{\sqrt 2}\, \cb
\left(\Yd{1}\right)_{ia}\, ,
\ \ \ 
\left( \wND \right)_{ia} = \frac{ve^{i\xi_1}}{\sqrt 2}\, \sb
\left(\Yd{1}\right)_{ia}\, ,
\end{equation}
and we obtain
\begin{equation}
\left( \wND \right)_{ia} = \tb\, \left( \wMD \right)_{ia}\, .
\label{eq:NtM}
\end{equation}
Now we use the \emph{Left conditions} in
eqs.~\eqref{eq:LeftProjection:00}-\eqref{P_i:MatrixForm}:
\begin{equation}
\left( \wND \right)_{ia} =
\ell_i^{[d]}\, \left( \wMD \right)_{ia}\, .
\label{eq:NellM}
\end{equation}
Combining eqs.~\eqref{eq:NtM} and \eqref{eq:NellM},
we find that
\begin{equation}
\left( \Yd{1} \right)_{i a} \neq 0
\ \ \Rightarrow \ \ \ell_i^{[d]} = \tb\, .
\label{eq:elli_tb}
\end{equation}

As a third step, we consider the possibility that
$\left( \Yd{2} \right)_{ib} \neq 0$.
A similar argument entails
\begin{equation}
\left( \Yd{2} \right)_{i b} \neq 0
\ \ \Rightarrow \ \ \ell_i^{[d]} = - \tbinv\, .
\label{eq:elli_-cotb}
\end{equation}
Comparing eq.~\eqref{eq:elli_tb} and \eqref{eq:elli_-cotb},
we conclude that the combination of an Abelian symmetry,
\textit{c.f.} eq.~\eqref{eq:Charges:00},
with the \emph{Left conditions} of eq.~\eqref{eq:LeftProjection:00}
implies that one cannot have simultaneously
$\left( \Yd{1} \right)_{ia} \neq 0$
and $\left( \Yd{2} \right)_{ib} \neq 0$,
for \emph{any} choices of $a$ and $b$.
So, for the \emph{Left condition},
$\Yd{1}$ and $\Yd{2}$ cannot both have nonzero matrix elements
in the same row.
This has the physical consequence that each doublet
$Q^0_{Li}$ couples to one and only one doublet $\Hd{k}$.

Moreover, we find the rule book for the assignment of $\ell_i^{[d]}$
in our models with \emph{Left conditions}:
\begin{eqnarray}
&&
\textrm{if } \left( \Yd{1} \right)_{ia} \textrm{ exists, then }
\ \ell_i^{[d]} = t_\beta\, ;
\nonumber\\
&&
\textrm{if } \left( \Yd{2} \right)_{ia} \textrm{ exists, then }
\ \ell_i^{[d]} = - t_\beta^{-1}\, .
\label{eq:RBook_Left}
\end{eqnarray}
One can easily see that the up sector matrices $\Yu{1}$ and $\Yu{2}$,
and the corresponding $\ell_i^{[u]}$ follow exactly the same rule.

\subsection{Left Models\label{sSEC:LEFT:Models}}
Omitting the trivial cases of type I or type II 2HDMs, for which the transformation properties in \refEQ{eq:Charges:00} have no flavour dependence (both $\LP{d}$ and $\LP{u}$ are in that case proportional to the identity matrix $\id$), we now address the different possible models which obey \emph{Left conditions}. 

\subsubsection{BGL models \label{mod:BGL}}
We start with the well known case of BGL models \cite{Branco:1996bq}.
The symmetry transformation is
\begin{equation}\label{eq:Lmodel:sym:1}
\Hd{2}\mapsto e^{i\theta}\Hd{2},\quad Q^0_{L3}\mapsto e^{-i\theta}Q^0_{L3},\quad d_{R3}^0\mapsto e^{-i2\theta}d_{R3}^0,\quad \theta\neq 0,\pi.
\end{equation}
The corresponding Yukawa coupling matrices are
\begin{equation}\label{eq:Lmodel:Y:1}
{\small 
\Yd{1}=\begin{pmatrix} \times & \times & 0\\  \times & \times & 0\\  0 & 0 & 0 \end{pmatrix}\!,
\Yd{2}=\begin{pmatrix} 0 & 0 & 0\\  0 & 0 & 0 \\  0 & 0 & \times\end{pmatrix}\!,
\Yu{1}=\begin{pmatrix} \times & \times & \times\\  \times & \times & \times \\  0 & 0 & 0 \end{pmatrix}\!,
\Yu{2}=\begin{pmatrix} 0 & 0 & 0\\  0 & 0 & 0 \\  \times & \times & \times\end{pmatrix}\!,
}
\end{equation}
where $\times$ denote arbitrary, independent, and (in general) non-vanishing
matrix entries.
Following the rule book in eq.~\eqref{eq:RBook_Left}
for the \emph{Left conditions},
we find immediately
\begin{equation}\label{eq:Lmodel:pro:1}
\wND=(\tb\PR{1}+\tb\PR{2}-\tbinv\PR{3})\,\wMD\,,\quad \wNU=(\tb\PR{1}+\tb\PR{2}-\tbinv\PR{3})\wMU\, .
\end{equation}
Here, the right-handed singlet transforming non-trivially in \refEQ{eq:Lmodel:sym:1}
is a down quark.
Such models are sometimes known as down-type BGL models, ``dBGL''.
In the particular implementation shown in \refEQ{eq:Lmodel:sym:1},
it is the third generation down quark which is involved;
this is known as a ``$b$ model''. 
We could equally well have substituted the
$d_{R3}^0\mapsto e^{-i2\theta}d_{R3}^0$ transformation in
\refEQ{eq:Lmodel:sym:1} by $d_{R1}^0\mapsto e^{-i2\theta}d_{R1}^0$,
or by $d_{R2}^0\mapsto e^{-i2\theta}d_{R2}^0$.
These are known as ``$d$ model'' and ``$s$ model'', respectively.
\\\underline{Parametrisation}\\
Following \refEQS{eq:Lmodel:pro:1} and \eqref{eq:LP:02}, one can write
\begin{equation}
\mND=(\tb\id-(\tti)\PRXL{3}{d})\mMD\,,\quad \mNU=(\tb\id-(\tti)\PRXL{3}{u})\mMU\,.
\end{equation}
Since $\Yd{1}$ and $\Yd{2}$ are block diagonal, $\wMD$ is block-diagonal too and then
\begin{equation}
\PRXL{3}{d}=\begin{pmatrix}0&0&0\\ 0&0&0\\ 0&0&1\end{pmatrix}\,.
\end{equation}
Using \refEQ{eq:LP:03},
\begin{equation}
\PRXL{3}{u}=\CKM\PRXL{3}{d}\CKMd,\quad\text{i.e.}\quad \left(\PRXL{3}{u}\right)_{ij}=\V{i3}\Vc{j3}=\V{ib}\Vc{jb}\,,
\end{equation}
and one obtains the final parametrisation for the physical couplings 
\begin{equation}\label{eq:Ndu:L1:00}
\left(\mND\right)_{ij}=
\delta_{ij}(\tb -(\tti)\delta_{j3})m_{d_j},\quad \left(\mNU\right)_{ij}=
(\tb\delta_{ij}-(\tti)\V{ib}\Vc{jb})m_{u_j}\,.
\end{equation}
Equation \eqref{eq:Ndu:L1:00} involves quarks masses, CKM mixings and $\tb$, but \emph{no new parameters}.
BGL models implement in a renormalizable 2HDM the ideas of Minimal Flavour Violation.
Besides this important property, BGL models are special in some respects that deserve comment: tree level FCNC are present either in the up or in the down quark sector, not in both (in this example, the $b$-dBGL model, they only appear in the up sector). The transformation properties in \refEQ{eq:Lmodel:sym:1} give a block diagonal form for the down Yukawa coupling matrices: this corresponds to the fact that some matrix conditions of the \emph{Right} type are also fulfilled for BGL models (this is not the case for the models in the next subsections). Finally, when the lepton sector is included in the picture, it was shown in \cite{Botella:2011ne} that the appropriate symmetry transformation group is $\ZZ{4}$, that is $\theta\to\pi/2$ in \refEQ{eq:Lmodel:sym:1}.

\subsubsection{Generalised BGL: gBGL \label{mod:gBGL}}
This second class of models is a generalisation of BGL models, introduced in \cite{Alves:2017xmk} (see also \cite{Joshipura:1990pi}); the defining transformation properties are
\begin{equation}\label{eq:Lmodel:sym:2}
\Hd{2}\mapsto -\Hd{2},\quad Q^0_{L3}\mapsto -Q^0_{L3},
\end{equation}
and the symmetry group is just  $\ZZ{2}$.
The corresponding Yukawa matrices are
\begin{equation}\label{eq:Lmodel:Y:2}
{\small 
\Yd{1}=\begin{pmatrix} \times & \times & \times\\ \times & \times & \times\\ 0 & 0 & 0 \end{pmatrix}\!,
\Yd{2}=\begin{pmatrix} 0 & 0 & 0\\ 0 & 0 & 0\\ \times & \times & \times \end{pmatrix}\!,
\Yu{1}=\begin{pmatrix} \times & \times & \times\\ \times & \times & \times\\ 0 & 0 & 0 \end{pmatrix}\!,
\Yu{2}=\begin{pmatrix} 0 & 0 & 0\\ 0 & 0 & 0\\ \times & \times & \times \end{pmatrix}\!.
}
\end{equation}
Following the rule book in eq.~\eqref{eq:RBook_Left}
for the \emph{Left conditions},
we find immediately
\begin{equation}\label{eq:Lmodel:pro:2}
\wND=(\tb\PR{1}+\tb\PR{2}-\tbinv\PR{3})\,\wMD\,,\quad \wNU=(\tb\PR{1}+\tb\PR{2}-\tbinv\PR{3})\wMU\,.
\end{equation}
%
\\\underline{Parametrisation}\\
While in the BGL model
(of section~\ref{mod:BGL})
$\Yd{1}$ and $\Yd{2}$ are block-diagonal, this is not the case here.
However, eqs.~\eqref{eq:Lmodel:pro:1} and \eqref{eq:Lmodel:pro:2} are identical\footnote{This is consistent with the fact that BGL can be recovered as a particular limit of generalised BGL models.}, giving again
\begin{equation}
\mND=(\tb\id-(\tti)\PRXL{3}{d})\mMD\,,\quad \mNU=(\tb\id-(\tti)\PRXL{3}{u})\mMU\,.
\end{equation}
Recalling \refEQ{eq:LP:02}, one can introduce complex unitary vectors
$\und{}$ and $\unu{}$ by
\begin{equation}
\und{j}\equiv
\left(\PR{3}\UdL \right)_{3j},\quad \unu{j}\equiv \left(\PR{3}\UuL\right)_{3j},
\end{equation}
in terms of which
\begin{equation}
\left(\PRXL{3}{d}\right)_{ij}=\undC{i}\und{j},\quad 
\left(\PRXL{3}{u}\right)_{ij}=\unuC{i}\unu{j}.
\end{equation}
The $\mND$ and $\mNU$ matrices are then given by
\begin{equation}\label{eq:Lmodel:Nq:fin:2}
\left(\mND\right)_{ij}=
(\tb \delta_{ij}-(\tti)\undC{i}\und{j})m_{d_j},\quad 
\left(\mNU\right)_{ij}=
(\tb\delta_{ij} -(\tti)\unuC{i}\unu{j})m_{u_j}\,.
\end{equation}
It is important to stress that $\undvec$ and $\unuvec$ are not independent.
From \refEQ{eq:LP:03},
\begin{equation}\label{eq:Lmodel:nund:2}
\unu{i}\V{ij}=\und{j}\,,
\end{equation}
and thus only four new independent parameters (besides quark masses, CKM mixings and $\tb$) appear in \refEQ{eq:Lmodel:Nq:fin:2}: two moduli, the third being fixed by normalization, and two relative phases, since the  products $\unqC{q}{i}\unq{q}{j}$ are insensitive to an overall phase.

\subsubsection{jBGL \label{mod:jBGL}}
The last case in this section is a new model
presented here for the first time
(see also \cite{Joshipura:1990xm}).
It is a sort of ``Flipped'' generalised BGL,
which follows from
\begin{equation}\label{eq:Lmodel:sym:3}
\Hd{2}\mapsto e^{i\theta}\Hd{2},\quad Q^0_{L3}\mapsto e^{-i\theta}Q^0_{L3},\quad d^0_{Rj}\mapsto e^{-i\theta}d^0_{Rj},\,j=1,2,3\,. 
\end{equation}
The corresponding Yukawa coupling matrices are
\begin{equation}\label{eq:Lmodel:Y:3}
{\small 
\Yd{1}=\begin{pmatrix} 0 & 0 & 0\\  0 & 0 & 0\\  \times & \times & \times \end{pmatrix}\!,
\Yd{2}=\begin{pmatrix} \times & \times & \times\\  \times & \times & \times\\  0 & 0 & 0 \end{pmatrix}\!,
\Yu{1}=\begin{pmatrix} \times & \times & \times\\  \times & \times & \times\\  0 & 0 & 0 \end{pmatrix}\!,
\Yu{2}=\begin{pmatrix} 0 & 0 & 0\\  0 & 0 & 0\\  \times & \times & \times \end{pmatrix}\!.
}
\end{equation}
The \emph{Left conditions} read in this case 
\begin{equation}\label{eq:Lmodel:pro:3}
\wND=(-\tbinv\PR{1}-\tbinv\PR{2}+\tb\PR{3})\,\wMD\,,\quad \wNU=(\tb\PR{1}+\tb\PR{2}-\tbinv\PR{3})\wMU\,.
\end{equation}
Notice how, with respect to \refEQ{eq:Lmodel:Y:2}, the structures of the down Yukawa matrices $\Yd{1}$ and $\Yd{2}$ are interchanged (while the $\Delta$ matrices remain the same).
\\\underline{Parametrisation}\\
Benefiting from the details given in the parametrisation of the gBGL models 
of section~\ref{mod:gBGL}, it is now straightforward to obtain
\begin{equation}\label{eq:Lmodel:Nq:fin:3}
\left(\mND\right)_{ij}=
(-\tbinv \delta_{ij}+(\tti)\undC{i}\und{j})m_{d_j},\quad
\left(\mNU\right)_{ij}=
(\tb\delta_{ij} -(\tti)\unuC{i}\unu{j})m_{u_j}\,,
\end{equation}
where, again, $\unu{i}\V{ij}=\und{j}$.
Notice the difference in the $\tb$ dependence of $\mND$ in
eq.~\eqref{eq:Lmodel:Nq:fin:3},
with respect to the gBGL case in eq.~\eqref{eq:Lmodel:Nq:fin:2}.

One can see that BGL is \emph{not} a particular case of jBGL.
Also, BGL is a particular limit of gBGL,
and jBGL is a sort of ``Flipped'' gBGL.
One might wonder whether there is some sort of
``Flipped'' BGL, \emph{obtainable from an Abelian symmetry},
which arises as a suitable limit of jBGL.
It is possible to see by inspection of the symmetry transformations
in eq.~\eqref{eq:Charges:00} that such a case is not allowed.

\subsection{Summary of models with \emph{Left conditions}\label{sSEC:LEFT:Summary}}
We summarize in Table~\ref{TAB:LeftModels} the main properties of the different models discussed in the previous subsections, which obey \emph{Left conditions}. For the BGL models of subsection \ref{mod:BGL} we display separately up and down type models (uBGL and dBGL respectively).
\begin{table}[h!tb]
\begin{center}
{\renewcommand{\arraystretch}{1.4}%
\begin{tabular}{|c|c|l|c|}
\hline
\backslashbox{{\tiny Model}}{{\tiny Properties}} & Sym. & Tree FCNC & Parameters \\
\hline
\multirow{2}{*}{G-W} & \multirow{2}{*}{$\ZZ{2}$} &
$\left(\mNU\right)_{ij}\propto\delta_{ij}m_{u_j}$ & \multirow{2}{*}{$\tb$, $m_{q_k}$}\\ 
 & & $\left(\mND\right)_{ij}\propto\delta_{ij}m_{d_j}$ &  \\ \hline
\multirow{2}{*}{uBGL ($t$)} & \multirow{2}{*}{$\ZZ{n\geq 4}$} & $\left(\mNU\right)_{ij}=\delta_{ij}(\tb -(\tti)\delta_{j3})m_{u_j}$ & \multirow{2}{*}{$\CKM$, $\tb$, $m_{q_k}$}\\ 
 & & $\left(\mND\right)_{ij}=(\tb\delta_{ij}-(\tti)\Vc{ti}\V{tj})m_{d_j}$ &  \\ \hline
\multirow{2}{*}{dBGL ($b$)} & \multirow{2}{*}{$\ZZ{n \geq 4}$} & $\left(\mNU\right)_{ij}=(\tb\delta_{ij}-(\tti)\V{ib}\Vc{jb})m_{u_j}$ & \multirow{2}{*}{$\CKM$, $\tb$, $m_{q_k}$}\\ 
 & & $\left(\mND\right)_{ij}=\delta_{ij}(\tb -(\tti)\delta_{j3})m_{d_j}$ &  \\ \hline
\multirow{2}{*}{gBGL} & \multirow{2}{*}{$\ZZ{2}$} & $\left(\mNU\right)_{ij}=(\tb\delta_{ij} -(\tti)\unuC{i}\unu{j})m_{u_j}$ & $\CKM$, $\tb$, $m_{q_k}$ \\ 
 & & $\left(\mND\right)_{ij}=(\tb \delta_{ij}-(\tti)\undC{i}\und{j})m_{d_j}$ &  $\unqvec{q}$(+4)\\ \hline
\multirow{2}{*}{jBGL} & \multirow{2}{*}{$\ZZ{n \geq 2}$} & $\left(\mNU\right)_{ij}=(\tb\delta_{ij} -(\tti)\unuC{i}\unu{j})m_{u_j}$ & $\CKM$, $\tb$, $m_{q_k}$\\ 
 & & $\left(\mND\right)_{ij}=(-\tbinv \delta_{ij}+(\tti)\undC{i}\und{j})m_{d_j}$ & $\unqvec{q}$(+4) \\ \hline
\end{tabular}
}
\caption{Models obeying the \emph{Left conditions} of
\refEQ{eq:LeftProjection:01}.
For the uBGL and dBGL models we only show the FCNC corresponding to one case, the \emph{top} and \emph{bottom} models, respectively. The first row shows Glashow-Weinberg models without tree level FCNC for comparison.\label{TAB:LeftModels}}
\end{center}
\end{table}
Since we have started from \emph{all} Abelian models consistent with
non-zero masses and a CKM matrix not block diagonal \cite{Ferreira:2010ir},
we are certain that 
Table~\ref{TAB:LeftModels} contains all models satisfying
the \emph{Left condition}.
We recovered the BGL \cite{Branco:1996bq} and gBGL \cite{Alves:2017xmk} models already
present in the literature, and proved
 that there exists only one
such new class of models, which we dubbed ``jBGL''.

\section{Symmetry Controlled Models with Right Conditions\label{SEC:RIGHT}}
In the previous section we have explored 2HDM whose symmetry under the Abelian transformations in \refEQ{eq:Charges:00} is supplemented by the requirement that the $\wMQ{q}$ and $\wNQ{q}$ obey the relations in \refEQ{eq:LeftProjection:00}, where $\wLP{q}$ in \refEQ{eq:LP:00} acts \emph{on the left}. In this section we analyse symmetry based models where we impose the conditions of \refEQ{eq:RightProjection:00}, $\wNQ{q}=\wMQ{q}\,\wRP{q}$, where $\wRP{q}$ in \refEQ{eq:RP:00} acts \emph{on the right}, that is, models which obey \emph{Right conditions}.
\subsection{Conditions in the mass basis\label{sSEC:RIGHT:Conditions}}
In the mass basis, \refEQ{eq:RightProjection:00} reads
\begin{equation}\label{eq:RightProjection:01}
\mND=\mMD\,\RP{d}\,,\quad \mNU=\mMU\RP{u}\,\,,
\end{equation}
with the transformed matrices
\begin{equation}
\RP{d}=\UdRd\,\wRP{d}\,\UdR,\quad \RP{u}=\UuRd\,\wRP{u}\,\UuR\,,
\end{equation}
and
\begin{equation}\label{eq:RP:01}
\RP{d}=r_1^{[d]}\PRXR{1}{d}+r_2^{[d]}\PRXR{2}{d}+r_3^{[d]}\PRXR{3}{d}\,,\quad
\RP{u}=r_1^{[u]}\PRXR{1}{u}+r_2^{[u]}\PRXR{2}{u}+r_3^{[u]}\PRXR{3}{u}\,.
\end{equation}
The transformed projection operators are now
\begin{equation}
\PRXR{j}{d}\equiv \UdRd\,\PR{j}\,\UdR,\quad\PRXR{j}{u}\equiv \UuRd\,\PR{j}\,\UuR\,.
\end{equation}
$\PRXR{j}{d}$ and $\PRXR{j}{u}$ are related via $\UuRd\UdR$, but, contrary to section \ref{sSEC:LEFT:Conditions}, this right-handed analog of the CKM matrix is completely arbitrary.
This straightforward yet crucial
difference among models with \emph{Left} and \emph{Right conditions}
will ultimately be responsible for the wider parametric freedom
of the latter.

\subsection{How to determine $r_i$\label{sSEC:RIGHT:r_i}}
Repeating the steps in section~\ref{sSEC:LEFT:ell_i},
one can easily establish here the following
rule book for the assignment of $r_i$
in our models with \emph{Right conditions}:
\begin{eqnarray}
&&
\textrm{if } \left( \Yd{1} \right)_{ai} \textrm{ exists, then }
\ r_i^{[d]} = \tb\, ;
\nonumber\\
&&
\textrm{if } \left( \Yd{2} \right)_{ai} \textrm{ exists, then }
\ r_i^{[d]} = - \tbinv\, .
\label{eq:RBook_Right}
\end{eqnarray}
One can also see that the up sector matrices $\Yu{1}$ and $\Yu{2}$,
and the corresponding $r_i^{[u]}$ follow exactly the same rule.

\subsection{Right Models\label{sSEC:RIGHT:Models}}
It is obvious that
cases in which both $\RP{d}$ and $\RP{u}$ are proportional to the identity
matrix have been discarded  automatically by the discussion of
models with \emph{Left conditions}.
But, for \emph{Right conditions} it is still possible to have
either $\RP{d}\propto\id$ or $\RP{u}\propto\id$ (but not both).
Among the six different types of models which obey \emph{Right conditions},
the first four have that property.

\subsubsection{Type A}
The first model follows from symmetry under
\begin{equation}\label{eq:Rmodel:sym:1}
\Hd{2}\mapsto e^{i\theta}\Hd{2},\quad u_{R3}^0\mapsto e^{i\theta}u_{R3}^0\, .
\end{equation}
The Yukawa coupling matrices in this case are
\begin{equation}\label{eq:Rmodel:Y:1}
{\small
\Yd{1}=\begin{pmatrix} \times & \times & \times\\  \times & \times & \times\\ \times & \times & \times \end{pmatrix}\!,
\Yd{2}=\begin{pmatrix} 0 & 0 & 0\\  0 & 0 & 0 \\  0 & 0 & 0\end{pmatrix}\!,
\Yu{1}=\begin{pmatrix} \times & \times & 0\\  \times & \times & 0 \\  \times & \times & 0 \end{pmatrix}\!,
\Yu{2}=\begin{pmatrix} 0 & 0 & \times\\  0 & 0 & \times \\  0 & 0 & \times\end{pmatrix}\!.
}
\end{equation}
We should mention that,
as explained in appendix~\ref{sSEC:APP:12},
it is immaterial whether $\Yu{1}$ contains the first two columns and
$\Yu{2}$ the third, or some other permutation is chosen.

Following the rule book in eq.~\eqref{eq:RBook_Right}
for the \emph{Right conditions},
we find immediately
\begin{equation}\label{eq:Rmodel:pro:1}
\wND=\wMD\,\tb\id,\quad \wNU=\wMU(\tb\PR{1}+\tb\PR{2}-\tbinv\PR{3})\,.
\end{equation}
%
\\\underline{Parametrisation}\\
Since only $\Yd{1}$ is non-zero, the down sector is trivial: $\mND=\tb\mMD$. For the up sector, however,
\begin{equation}
\mNU=\mMU(\tb\id-(\tti)\PRXR{3}{u})\,.
\end{equation}
Similarly to the models in section \ref{SEC:LEFT}, one can introduce a complex unitary vector $\rnuvec$
\begin{equation}
\rnu{j}\equiv
\left(\PR{3}\UuR\right)_{3j},
\end{equation}
in terms of which
\begin{equation}
\left(\PRXR{3}{u}\right)_{ij}=\rnuC{i}\rnu{j},
\end{equation}
and thus
\begin{equation}\label{eq:Rmodel:Nq:fin:1}
\left(\mND\right)_{ij}=m_{d_i} \tb \delta_{ij},\quad
\left(\mNU\right)_{ij}=m_{u_i}(\tb\delta_{ij} -(\tti)\rnuC{i}\rnu{j})\,.
\end{equation}
Therefore, besides quark masses and $\tb$, only four new independent parameters appear in \refEQ{eq:Rmodel:Nq:fin:1}.

\subsubsection{Type B}
The second model follows from the symmetry
\begin{equation}\label{eq:Rmodel:sym:2}
\Hd{2}\mapsto e^{i\theta}\Hd{2},
\quad u_{R1}^0\mapsto e^{i\theta}u_{R1}^0,
\quad u_{R2}^0\mapsto e^{i\theta}u_{R2}^0.
\end{equation}
The corresponding Yukawa coupling matrices are
\begin{equation}\label{eq:Rmodel:Y:2}
{\small
\Yd{1}=\begin{pmatrix} \times & \times & \times\\  \times & \times & \times\\
\times & \times & \times \end{pmatrix}\!,
\Yd{2}=\begin{pmatrix} 0 & 0 & 0\\  0 & 0 & 0 \\  0 & 0 & 0\end{pmatrix}\!,
\Yu{1}=\begin{pmatrix} 0 & 0 & \times\\  0 & 0 & \times \\  0 & 0 & \times \end{pmatrix}\!,
\Yu{2}=\begin{pmatrix} \times & \times & 0\\  \times & \times & 0\\
\times & \times & 0\end{pmatrix}\!.
}
\end{equation}
Notice how, with respect to the previous model
in \refEQ{eq:Rmodel:Y:1},
the forms of $\Yu{1}$ and $\Yu{2}$ are interchanged in \refEQ{eq:Rmodel:Y:2}.
Thus, our Type B model is a sort of Flipped Type A model.
The \emph{Right conditions} become
\begin{equation}\label{eq:Rmodel:pro:2}
\wND=\wMD\,\tb\id,\quad \wNU=\wMU(-\tbinv\PR{1}-\tbinv\PR{2}+\tb\PR{3})\,.
\end{equation}
%
\\\underline{Parametrisation}\\
Given the parametrisation of the previous case, it follows immediately that in this case:
\begin{equation}\label{eq:Rmodel:Nq:fin:2}
\left(\mND\right)_{ij}=
m_{d_i} \tb \delta_{ij},\quad
\left(\mNU\right)_{ij}=
m_{u_i}(-\tbinv\delta_{ij} +(\tti)\rnuC{i}\rnu{j})\,.
\end{equation}
Notice the different $\tb$ dependence in \refEQ{eq:Rmodel:Nq:fin:2} with respect to \refEQ{eq:Rmodel:Nq:fin:1}.

\subsubsection{Type C}
The transformation properties for this model are
\begin{equation}\label{eq:Rmodel:sym:C}
\Hd{2}\mapsto e^{i\theta}\Hd{2},\quad d_{R3}^0\mapsto e^{-i\theta}d_{R3}^0,
\end{equation}
and the \emph{Right conditions} read
\begin{equation}\label{eq:Rmodel:pro:C}
\wND=\wMD(\tb\PR{1}+\tb\PR{2}-\tbinv\PR{3}),\quad \wNU=\wMU\,\tb\id\,.
\end{equation}
The Yukawa coupling matrices are in this case
\begin{equation}\label{eq:Rmodel:Y:C}
{\small
	\Yd{1}=\begin{pmatrix} \times & \times & 0\\  \times & \times & 0 \\  \times & \times & 0 \end{pmatrix}\!,
	\Yd{2}=\begin{pmatrix} 0 & 0 & \times\\  0 & 0 & \times \\  0 & 0 & \times\end{pmatrix}\!,
	\Yu{1}=\begin{pmatrix} \times & \times & \times\\  \times & \times & \times\\ \times & \times & \times \end{pmatrix}\!,
	\Yu{2}=\begin{pmatrix} 0 & 0 & 0\\  0 & 0 & 0 \\  0 & 0 & 0\end{pmatrix}\!.
}
\end{equation}
\\\underline{Parametrisation}\\
Since
\begin{equation}
\mND=\mMD(\tb\id-(\tti)\PRXR{3}{d}),\quad \mNU=\tb\mMU\, ,
\end{equation}
defining
\begin{equation}
\rnd{j}\equiv \left(\PR{3}\UdR\right)_{3j},
\end{equation}
and
\begin{equation}
\left(\PRXR{3}{d}\right)_{ij}=\rndC{i}\rnd{j},
\end{equation}
we find
\begin{equation}\label{eq:Rmodel:Nq:fin:C}
\left(\mND\right)_{ij}=m_{d_i}(\tb\delta_{ij} -(\tti)\rndC{i}\rnd{j}),\quad
\left(\mNU\right)_{ij}=\tb m_{u_i}\delta_{ij}\,,
\end{equation}
implying that, besides quark masses and $\tb$, only four new independent parameters appear in \refEQ{eq:Rmodel:Nq:fin:C}.
A particular case of these models appears in ref.~\cite{Lavoura:1994ty},
with all coefficients taken as real in order to have an exclusive spontaneous
origin for CP violation (no CKM CP violation).
As such,
there are in ref.~\cite{Lavoura:1994ty} only two instead of four parameters arising from $\rnd{j}$.

\subsubsection{Type D}
The transformation properties for this model are
\begin{equation}\label{eq:Rmodel:sym:D}
\Hd{2}\mapsto e^{i\theta}\Hd{2},\quad d_{R1}^0\mapsto e^{-i\theta}d_{R1}^0,\quad d_{R2}^0\mapsto e^{-i\theta}d_{R2}^0,
\end{equation}
and the \emph{Right conditions} read
\begin{equation}\label{eq:Rmodel:pro:D}
\wND=\wMD(-\tbinv\PR{1}-\tbinv\PR{2}+\tb\PR{3}),\quad \wNU=\wMU\,\tb\id\,.
\end{equation}
The Yukawa coupling matrices are in this case
\begin{equation}\label{eq:Rmodel:Y:D}
{\small
\Yd{1}=\begin{pmatrix} 0 & 0 & \times\\  0 & 0 & \times \\  0 & 0 & \times\end{pmatrix}\!,
\Yd{2}=\begin{pmatrix} \times & \times & 0\\  \times & \times & 0 \\  \times & \times & 0 \end{pmatrix}\!,
\Yu{1}=\begin{pmatrix} \times & \times & \times\\  \times & \times & \times\\ \times & \times & \times \end{pmatrix}\!,
\Yu{2}=\begin{pmatrix} 0 & 0 & 0\\  0 & 0 & 0 \\  0 & 0 & 0\end{pmatrix}\!.
}
\end{equation}
\\\underline{Parametrisation}\\
Here
\begin{equation}
\mND=\mMD(-\tbinv\id+(\tti)\PRXR{3}{d}),\quad \mNU=\tb\mMU\, ,
\end{equation}
from which
\begin{equation}\label{eq:Rmodel:Nq:fin:D}
\left( \mND \right)_{ij}=m_{d_i}(-\tb\delta_{ij} +(\tti)\rndC{i}\rnd{j}),\quad
\left( \mNU \right)_{ij}=\tb m_{u_i}\delta_{ij}\,.
\end{equation}
Therefore, besides quark masses and $\tb$, only four new independent parameters appear in \refEQ{eq:Rmodel:Nq:fin:D}.
\subsubsection{Type E}
The transformation properties for this model are
\begin{equation}\label{eq:Rmodel:sym:3}
\Hd{2}\mapsto e^{i\theta}\Hd{2},\quad d_{R3}^0\mapsto e^{-i\theta}d_{R3}^0,\quad u_{R3}^0\mapsto e^{i\theta}u_{R3}^0,
\end{equation}
and the corresponding Yukawa coupling matrices are
\begin{equation}\label{eq:Rmodel:Y:3}
{\small
\Yd{1}=\begin{pmatrix} \times & \times & 0\\  \times & \times & 0\\ \times & \times & 0 \end{pmatrix}\!,
\Yd{2}=\begin{pmatrix} 0 & 0 & \times\\  0 & 0 & \times \\  0 & 0 & \times\end{pmatrix}\!,
\Yu{1}=\begin{pmatrix} \times & \times & 0\\  \times & \times & 0 \\  \times & \times & 0 \end{pmatrix}\!,
\Yu{2}=\begin{pmatrix} 0 & 0 & \times\\  0 & 0 & \times \\  0 & 0 & \times\end{pmatrix}\!,
}
\end{equation}
leading to the \emph{Right conditions}
\begin{equation}\label{eq:Rmodel:pro:3}
\wND=\wMD(\tb\PR{1}+\tb\PR{2}-\tbinv\PR{3}),\quad \wNU=\wMU(\tb\PR{1}+\tb\PR{2}-\tbinv\PR{3})\,.
\end{equation}
\\\underline{Parametrisation}\\
While in the previous models one quark sector had a trivial structure (since $\Gamma_2=0$ in types A and B, while $\Delta_2=0$ in types C and D), that is not the case in \refEQ{eq:Rmodel:Y:3}, and one naturally expects an increase in the number of parameters. An appropriate parametrisation is obtained along the same lines as before. With
\begin{equation}
\mND=\mMD(\tb\id-(\tti)\PRXR{3}{d})\,,\quad\mNU=\mMU(\tb\id-(\tti)\PRXR{3}{u})\,,
\end{equation}
but \emph{two} complex unitary vectors are now necessary, $\rndvec$ and $\rnuvec$,
defined by
\begin{equation}
\rnd{j}\equiv \left(\PR{3}\UdR\right)_{3j},\quad
\rnu{j}\equiv \left(\PR{3}\UuR\right)_{3j},
\end{equation}
and in terms of which
\begin{equation}
\left(\PRXR{3}{d}\right)_{ij}=\rndC{i}\rnd{j},\quad
\left(\PRXR{3}{u}\right)_{ij}=\rnuC{i}\rnu{j}.
\end{equation}
The parametrisation of this model is then
\begin{equation}\label{eq:Rmodel:Nq:fin:3}
\left(\mND\right)_{ij}=
m_{d_i} (\tb\delta_{ij} -(\tti)\rndC{i}\rnd{j}),\quad
\left(\mNU\right)_{ij}=
m_{u_i} (\tb\delta_{ij} -(\tti)\rnuC{i}\rnu{j})\,.
\end{equation}
It is important to notice that now, besides the quark masses and $\tb$, four new independent real parameters enter \refEQ{eq:Rmodel:Nq:fin:3} via $\rnd{j}$ and another four via $\rnu{j}$. Contrary to the situation in models with \emph{Left conditions} in section \ref{SEC:LEFT}, where the CKM matrix ties $\unuvec$ and $\undvec$, and it is fixed or given by another sector of the complete model (the couplings of quarks to the $W$ gauge boson), in models with \emph{Right conditions} there is no analog of the CKM matrix to connect $\rnuvec$ and $\rndvec$ in a fixed manner\footnote{Interpreting the situation the other way around, \refEQ{eq:Rmodel:Nq:fin:3} would provide a window of sensitivity to the right-handed analog of CKM (for example, in extensions to models with a gauged
$SU(2)_L \otimes SU(2)_R$ symmetry).}.

\subsubsection{Type F}
The transformation properties of this last model are
\begin{equation}\label{eq:Rmodel:sym:4}
\Hd{2}\mapsto e^{i\theta}\Hd{2},\quad d_{R3}^0\mapsto e^{-i\theta}d_{R3}^0,\quad u_{R1}^0\mapsto e^{i\theta}u_{R1}^0,\quad u_{R2}^0\mapsto e^{i\theta}u_{R2}^0,
\end{equation}
and the corresponding Yukawa coupling matrices have the following form:
\begin{equation}\label{eq:Rmodel:Y:4}
{\small
\Yd{1}=\begin{pmatrix} \times & \times & 0\\  \times & \times & 0\\ \times & \times & 0 \end{pmatrix}\!,
\Yd{2}=\begin{pmatrix} 0 & 0 & \times\\  0 & 0 & \times \\  0 & 0 & \times\end{pmatrix}\!,
\Yu{1}=\begin{pmatrix} 0 & 0 & \times\\  0 & 0 & \times \\  0 & 0 & \times \end{pmatrix}\!,
\Yu{2}=\begin{pmatrix} \times & \times & 0\\  \times & \times & 0\\ \times & \times & 0 \end{pmatrix}\!.
}
\end{equation}
Notice how, with respect to the previous model in \refEQ{eq:Rmodel:Y:3},
the forms of $\Yu{1}$ and $\Yu{2}$ are interchanged in \refEQ{eq:Rmodel:Y:4}.
Thus, our Type F model is a sort of Flipped Type E model.
The \emph{Right conditions} become
\begin{equation}\label{eq:Rmodel:pro:4}
\wND=\wMD(\tb\PR{1}+\tb\PR{2}-\tbinv\PR{3}),\quad \wNU=\wMU(-\tbinv\PR{1}-\tbinv\PR{2}+\tb\PR{3})\,.
\end{equation}
\\\underline{Parametrisation}\\
Parametrising this last model follows trivially from the previous one:
\begin{equation}\label{eq:Rmodel:Nq:fin:4}
\left(\mND\right)_{ij}=
m_{d_i} (\tb\delta_{ij} -(\tti)\rndC{i}\rnd{j}),\quad
\left(\mNU\right)_{ij}=
m_{u_i} (-\tbinv\delta_{ij}+(\tti)\rnuC{i}\rnu{j})\,.
\end{equation}
The same comments made in Type E apply to the parameter count in Type F models:
besides the quark masses and $\tb$, as in \refEQ{eq:Rmodel:Nq:fin:3},
four new independent real parameters enter
\refEQ{eq:Rmodel:Nq:fin:4} via $\rnd{j}$ and another four via $\rnu{j}$.
\subsection{Summary of models with \emph{Right conditions}\label{sSEC:Right:Summary}}
We summarize in Table \ref{TAB:RightModels} the main properties of the different models discussed in the previous subsections, which obey \emph{Right conditions}.
\begin{table}[h!tb]
\begin{center}
{\renewcommand{\arraystretch}{1.4}%
\begin{tabular}{|c|c|l|c|}
\hline
\backslashbox{{\tiny Model}}{{\tiny Properties}} & Sym. & Tree FCNC & Parameters \\ \hline
\multirow{2}{*}{G-W} & \multirow{2}{*}{$\ZZ{2}$} & $\left(\mNU\right)_{ij}\propto m_{u_i} \delta_{ij}$ & \multirow{2}{*}{$\tb$, $m_{q_k}$}\\
 & & $\left(\mND\right)_{ij}\propto m_{d_i} \delta_{ij}$ & \\ \hline
\multirow{2}{*}{Type A} & \multirow{2}{*}{$\ZZ{n\geq 2}$} & 			$\left(\mNU\right)_{ij}= m_{u_i} (\tb\delta_{ij} -(\tti)\rnuC{i}\rnu{j})$
 & $\tb$, $m_{q_k}$\\
 & & $\left(\mND\right)_{ij}= m_{d_i} \tb \delta_{ij}$ & $\rnuvec$(+4) \\ \hline	
\multirow{2}{*}{Type B} & \multirow{2}{*}{$\ZZ{n\geq 2}$} & 			$\left(\mNU\right)_{ij}= m_{u_i} (-\tbinv\delta_{ij}+(\tti)\rnuC{i}\rnu{j})$ & $\tb$, $m_{q_k}$\\
 & & $\left(\mND\right)_{ij}= m_{d_i} \tb \delta_{ij}$ & $\rnuvec$(+4) \\
\hline
\multirow{2}{*}{Type C} & \multirow{2}{*}{$\ZZ{n\geq 2}$} & $\left(\mNU\right)_{ij}= m_{u_i} \tb \delta_{ij}$ & $\tb$, $m_{q_k}$\\
 & & $\left(\mND\right)_{ij}= m_{d_i} (\tb\delta_{ij}-(\tti)\rndC{i}\rnd{j})$ & $\rndvec$(+4)\\ \hline
\multirow{2}{*}{Type D} & \multirow{2}{*}{$\ZZ{n\geq 2}$} & $\left(\mNU\right)_{ij}= m_{u_i} \tb \delta_{ij}$ & $\tb$, $m_{q_k}$\\
 & &  $\left(\mND\right)_{ij}= m_{d_i} (-\tbinv\delta_{ij}+(\tti)\rndC{i}\rnd{j})$ & $\rndvec$(+4)\\ \hline
\multirow{2}{*}{Type E} & \multirow{2}{*}{$\ZZ{n\geq 2}$} &
$\left(\mNU\right)_{ij}= m_{u_i} (\tb\delta_{ij} -(\tti)\rnuC{i}\rnu{j})$
 & $\tb$, $m_{q_k}$\\
 & & $\left(\mND\right)_{ij}= m_{d_i} (\tb\delta_{ij}-(\tti)\rndC{i}\rnd{j})$ & $\rnuvec$,$\rndvec$(+8) \\ \hline
\multirow{2}{*}{Type F} & \multirow{2}{*}{$\ZZ{n\geq 2}$} &			$\left(\mNU\right)_{ij}= m_{u_i}(-\tbinv\delta_{ij}+(\tti)\rnuC{i}\rnu{j})$ & $\tb$, $m_{q_k}$ \\
 & & $\left(\mND\right)_{ij}= m_{d_i} (\tb\delta_{ij}-(\tti)\rndC{i}\rnd{j})$ & $\rnuvec$,$\rndvec$(+8) \\ \hline
\end{tabular}
}
\caption{Models obeying \emph{Right conditions}, \refEQ{eq:RightProjection:01}. As in Table \ref{TAB:LeftModels}, the first row shows Glashow-Weinberg models without tree level FCNC for comparison.\label{TAB:RightModels}}
\end{center}
\end{table}



\section{Phenomenology\label{SEC:PHENO}}
In the previous sections we have presented different classes of models which include controlled tree-level FCNC; different cases within the same class share the same number of parameters, and this number varies among different classes. This section is devoted to a discussion of aspects related to the phenomenology of the different models.

Eq.~\eqref{eq:Yukawa:01} shows the relevant Lagrangian.
We can read from it directly the couplings of the charged scalars,
involving $V N_d$, $V^\dagger N_u$, and the Hermitian conjugates
$N_d^\dagger V^\dagger$ and $N_u^\dagger V$:
%
\begin{equation}
-\frac{v}{\sqrt 2}\mathscr L_{\rm Y}
\supset
H^+\ \bar{u}_\alpha
\left[
\left( V N_d\right)_{\alpha k} \gamma_R
- \left( N_u^\dagger V\right)_{\alpha k} \gamma_L
\right]\, d_k + \text{H.c.},
\label{eq:h+lagrang}
\end{equation}
where $\gamma_{R,L} = (1 \pm \gamma_5)/2$,
and sums over the up quark (down quark) index $\alpha$ ($k$)
are implicit.
To find the couplings with the neutral scalars, one must specify
the scalar potential.
In models with a $\ZZ{2}$ symmetry softly broken,
one can have CP violation in the scalar sector,
spontaneous \cite{Branco:1985aq} or explicit --
for recent reviews, see for example
\cite{Fontes:2014xva, Grzadkowski:2016szj, Fontes:2017zfn}.
Conversely,
if CP is conserved, then $I^0$ in eq.~\eqref{eq:HiggsBasis:02}
is a CP-odd mass eigenstate, usually denoted by $A$.
Still,
the scalars $h^0$ and $R^0$ written in
the Higgs basis of eq.~\eqref{eq:HiggsBasis:02} mix into the
mass eigenstate basis of CP-even neutral scalars $h$ and $H$
via an angle $\beta-\alpha$.
As a result,
the couplings of these scalars become of the type
\begin{eqnarray}
&&
s_{\beta-\alpha} M_q + c_{\beta-\alpha} N_q\, ,
\nonumber\\
&&
- c_{\beta-\alpha} M_q + s_{\beta-\alpha} N_q\, ,
\label{eq:h_H_couplings}
\end{eqnarray}
for the lighter $h$ and heavier $H$ scalars,
respectively ($\cos x\equiv c_x$, $\sin x\equiv s_x$).
We know from the decays of the 125 GeV scalar \cite{Khachatryan:2016vau} that
$s_{\beta-\alpha}$ should lie close to 1.
Besides the $\beta-\alpha$ mixing effect present in
the usual type I and type II (and X and Y),
we see that there are now FCNC controlled by $N_q$,
even for the 125 GeV scalar (which we take to be the lighter state $h$).
These effects are $c_{\beta-\alpha}$ suppressed in $h$,
but not in $H$ (or the charged scalars $H^\pm$).
The effects of $N_q$ can appear in both flavour changing
and in flavour conserving couplings.
The former require a non-diagonal $N_q$,
while the latter exist even if $N_q$ turned out to be diagonal.
An important result of our paper is that for the models discussed here,
non diagonal couplings, when they exist,
are in every case proportional to
\begin{equation}
\pm (\tti)\unqC{q}{i}\unqC{q}{j}m_{q_j}\,,\quad \, \pm m_{q_i}(\tti)\rnqC{q}{i}\rnqC{q}{j}\,,
\label{eq:FCNC}
\end{equation}
for \emph{Left} and \emph{Right} models, respectively.
Since $(\tti)=2/s_{2\beta}$ is equal or larger than 2,
and could in principle be arbitrarily large,
this could overcome the $c_{\beta-\alpha}$ suppression
of FCNC for the 125 GeV scalar mentioned above.

In short,
there are two obvious ingredients of these models:
there are new scalar particles, charged and neutral;
and there are FCNC at the tree level.
Starting from them, the possible New Physics clues
motivating interest in these 2HDMs can be classified in
\begin{itemize}
\item deviations from SM expectations in the flavour conserving processes involving the 125 GeV Higgs-like scalar,
\item possible sizable FCNC processes involving the 125 GeV Higgs-like scalar,
\item proposed searches for new fundamental scalars.
\end{itemize}
The division is to some extent arbitrary since all three aspects are related: through mixing in the scalar sector, the 125 GeV Higgs-like scalar inherits tree-level FCNC and modified flavour conserving couplings. With those eventual clues, one can then ask two different questions:
\begin{enumerate}
\item how can one fix or extract parameters of a given model?
\item how can one tell apart different models?
\end{enumerate}

The BGL models of section \ref{mod:BGL} have already been extensively studied \cite{Botella:2009pq,Botella:2011ne},
including phenomenological aspects
\cite{Botella:2014ska,Bhattacharyya:2014nja,Botella:2015hoa},
while the gBGL models of section \ref{mod:gBGL} were introduced in \cite{Alves:2017xmk}, including some insight into their phenomenology.
All other models can be implemented via a $\ZZ{2}$ symmetry,
which we consider softly broken.
As a result,
there is a decoupling limit and all SM predictions can be recovered
by taking the extra scalars very massive.
Conversely,
as one makes the scalars lighter,
the matrices $\mND$ and $\mNU$
(and their effect on both flavour changing and flavour conserving couplings)
become more important.
The crucial result in eq.~\eqref{eq:FCNC} means that the phenomenological
analysis is very similar in all cases,
and follows the same steps discussed for the gBGL models
in ref.~\cite{Alves:2017xmk}.

From eqs.~\eqref{eq:h_H_couplings}
and \eqref{eq:FCNC} it turns out that the 125 GeV Higgs
has flavour changing Yukawa coupling typically of the form
\begin{equation}
-\mathscr L_{h \bar{q}_{i}q_{j}}
=
-h\bar{q}_{i}Y_{ij}q_{j}\, ,
\end{equation}
where
\begin{equation}
Y_{ij} =
c_{\beta -\alpha }\left( \tti\right)\unqC{q}{i}\unq{q}{j}\frac{m_{q_j}}{v}\,\quad \text{or}\quad
Y_{ij} =
c_{\beta -\alpha }\left( \tti\right)\frac{m_{q_i}}{v}\rnqC{q}{i}\rnq{q}{j}\,.
\end{equation}
These couplings, appearing in all the models, contribute to the $\Delta F=2$
neutral mesons mixing amplitude.
In all the new models present in this paper we have an arbitrary
complex unitary vector $\hat {\rm v}=\unqvec{q},\rnqvec{q}$, and, therefore,
the maximum intensity of these flavour changing coupling can be reached when
$\abs{\hat {\rm v}_i\hat {\rm v}_j}$ takes its maximum value of 1/2.
It is under this assumption that, by imposing the constraints
from
$K^{0}$--$\bar{K}^{0}$, $B_{d}^{0}$--$\bar{B}_{d}^{0}$, 
$B_{s}^{0}$--$\bar{B}_{s}^{0}$, and $D^{0}$--$\bar{D}^{0}$,
we can get an universal bound for
$\abs{c_{\beta -\alpha }(\tb+\tbinv)}$.
Following the analysis of ref.~\cite{Botella:2015hoa} and the
constraints in ref.~\cite{Blankenburg:2012ex},
one can conclude that all the models presented here are safe over the entire
parameter space,
provided we take
\begin{equation}
\ABS{c_{\beta -\alpha }(\tti)} \leq 0.02\ .
\end{equation}
It has to be stressed that in a large region of the
$\hat{\rm v}$ parameter space
$\abs{c_{\beta -\alpha}(\tti)}$
can span almost all the theoretically allowed
parameter region, even reaching values of order one.
In this paper we will not consider semileptonic
$\Delta F=1$ processes because any constraint
will introduce extra model dependences coming from the
specific choice one might make for the leptonic sector.

In general,
$Y_{ij}$ presents an extremely important $m_{q_i}/v$ suppression,
except in the case where $q_i=t$ corresponds to the top quark.
In those models with FCNC in the up sector one must also check the constraints
arising from rare top decays,
such as $t\to hc$ and $t\to hu$.
One finds \cite{Alves:2017xmk}
\begin{equation}
\text{Br}\left(t\to hq\right) = 0.13 \ABS{\frac{\hat{\rm v}_{q}\hat{\rm v}_{t}}{V_{tq}}}^{2}\ABS{c_{\beta -\alpha }(\tti)}^{2}\, .
\end{equation}
Taking into account the experimental bounds
from ATLAS \cite{Aad:2014dya, Aaboud:2017mfd}
and CMS \cite{Khachatryan:2014jya, Khachatryan:2016atv},
we get
\begin{equation}
\ABS{c_{\beta -\alpha }(\tti)} \leq 0.4\, ,
\end{equation}
again for the maximum theoretical value $\abs{\hat{\rm v}_{i}\hat{\rm v}_{j}}=1/2$.

A full parameter scan lies beyond the scope of this work, but we concentrate here on an important aspect. Let us imagine that the new particles and FCNC effects discussed in this article had been detected. In that case, what properties could be relevant to tell apart different models?\footnote{It goes without saying that the more experimental signals are available, the better the identification can be established. On that respect, available information on Higgs production $\times$ decay signal strenghts can provide some sensitivity to the diagonal couplings in \refEQ{eq:h_H_couplings}. However, that avenue is not as distinctive as tree level FCNC.}
\begin{itemize}
\item uBGL, dBGL and types A,B,C and D only have tree level FCNC in one quark sector, up or down, not both. Furthermore, uBGL and dBGL are fixed in terms of the CKM matrix, while the couplings $\rnq{q}{j}$ in A, B, C and D are free parameters.
\item gBGL, jBGL and types E and F have tree level FCNC in both sectors. However, in gBGL and jBGL, the parameters controlling them, $\unu{j}$ and $\und{j}$, are not independent, they are related through CKM, \refEQ{eq:Lmodel:nund:2}, while that is not the case in models E and F where $\rnu{j}$ and $\rnd{j}$ are independent. 
\end{itemize}
With these considerations, it is clear that FCNC allow for some discrimination among models, but it is not complete. Consider for example gBGL and jBGL models. $\mNU$ has the same structure in both cases, and so do the off diagonal couplings of neutral scalars with down quarks\footnote{The sign is irrelevant when considering exclusively off diagonal couplings with neutral scalars.}. A similar comment applies to models A vs. B, C vs. D and E vs. F. The relevant question is then: can one tell apart gBGL from jBGL? (And similarly A from B, C from D and E from F) 
It is interesting to consider the decays of the charged Higgs into quarks.
We find
\begin{equation}
\Gamma(H^+ \to \bar{u}_\alpha d_k)
=
\frac{N_c m_{H^\pm}}{8 \pi}
\sqrt{\beta_+ \beta_-}
\left[
\left| a_{\alpha k}\right|^2 \beta_+
+ \left| b_{\alpha k}\right|^2 \beta_-
\right]\, ,
\label{eq:decay_rate}
\end{equation}
where
\begin{equation}
\beta_\pm = 1 - \left(\frac{m_{u_\alpha} \pm m_{d_k}}{m_{H^\pm}} \right)^2\, ,
\end{equation}
and
\begin{eqnarray}
a_{\alpha k} &=&
\frac{1}{\sqrt{2} v} \left[
\left(\CKM \mND \right)_{\alpha k}
- \left(\mNUd \CKM \right)_{\alpha k}
\right]\, ,
\nonumber\\
b_{\alpha k} &=&
\frac{-i}{\sqrt{2} v} \left[
\left(\CKM \mND \right)_{\alpha k}
+ \left(\mNUd \CKM \right)_{\alpha k}
\right]\, ,
\label{eq:a_and_b}
\end{eqnarray}
are the scalar and pseudoscalar couplings one finds when rewriting the
Lagrangian in eq.~\eqref{eq:h+lagrang} as
\begin{equation}
-\mathscr L_{\rm Y}
\supset
H^+\ \bar{u}_\alpha
\left[ a_{\alpha k} + i\,  b_{\alpha k} \gamma_5\right] d_k + \text{H.c.}
\end{equation}
The $i$ in the second equation \eqref{eq:a_and_b} is crucial.
In a reasoning similar to that used in ref.~\cite{Nebot:2015wsa} for
flavour changing neutral couplings, one can prove that
\begin{equation}
\textrm{Re} \left( a_{\alpha k} b_{\alpha k}^\ast\right)\neq 0
\quad\Rightarrow\quad
\textrm{CP violation}.
\end{equation}
In our case
\begin{equation}
\textrm{Re} \left( a_{\alpha k} b_{\alpha k}^\ast\right)
\propto
\textrm{Re}
\left\{
i
\left[
\left( \CKM \mND\right)^2_{\alpha k}
-
\left( \mNDd \CKM \right)^2_{\alpha k}
\right]
\right\},
\end{equation}
showing that CP is conserved if $\mND$, $\mNU$, and $\CKM$ are real,
as expected\footnote{Naturally, a tree level calculation
would yield $\Gamma(H^+ \to \bar{u}_\alpha d_k) =
\Gamma(H^- \to u_\alpha \bar{d}_k)$
even for complex parameters,
consistent with the absense of direct CP violation in the presence
of a single diagram. Including some loop diagram(s), with different
strong and weak phases, would yield a CP violating asymmetry.}.
We can now see that the decays in
eqs.~\eqref{eq:decay_rate}-\eqref{eq:a_and_b}
provides access to the beating of $\mND,\mNU$ against the CKM matrix $V$,
thus permitting a distinction between gBGL and jBGL models.


\section{Conclusions\label{SEC:CONC}}

The recent discovery of a scalar particle prompted the search for
more scalars and re-spurring the study of models with two Higgs doublets.
A general two Higgs doublet model has double the number of Yukawa couplings
already present in the SM. It would seem that this would lead us even farther
away from an understanding of the flavour sector.
Moreover,
2HDM typically lead to FCNC, which are tightly constrained by experiment.
In this article we entertain the possibility that these two issues
are solved in a natural way by the presence of Abelian symmetries.
We are inspired by the BGL \cite{Branco:1996bq} models,
where FCNC are entirely determined by the CKM matrix elements,
and by gBGL \cite{Alves:2017xmk} models,
which have a larger parametric freedom.

We show that such models can be obtained by enhancing an Abelian
symmetry with the \emph{Left condition} in eq.~\eqref{eq:LeftProjection:00}.
Since Ferreira and Silva \cite{Ferreira:2010ir}
had listed all 2HDM models constrained by
an Abelian symmetry and consistent with nonzero quark masses and
a non-diagonal CKM,
we could perform an exhaustive search for all such models,
and show there is one, and only one, further class of models obeying
the \emph{Left condition}, which we dubbed jBGL.

We have developed a similar \emph{Right condition}
\eqref{eq:RightProjection:00} and again performed an exhaustive search
over the set of models with an Abelian symmetry.
We identified six new classes of models,
named Types A through F.
For all cases,
the FCNC matrices $\mND$ and $\mNU$ have been written in terms
of masses, $\tan{\beta}$, CKM entries, and
vectors containing all the remaining parametric freedom.
All FCNC couplings have the generic form in eq.~\eqref{eq:FCNC}.
Finally, we discussed how one could in principle tell these models apart,
by concentrating on the use of charged Higgs decays to disentangle
gBGL from jBGL models

\section*{Acknowledgments}
This work is partially supported by Spanish MINECO under grant FPA2015-68318-R, FPA2017-85140-C3-3-P and by the Severo Ochoa Excellence Center Project SEV-2014-0398, by Generalitat Valenciana under grant GVPROMETEOII 2014-049 and by Funda\c{c}\~ao para a Ci\^encia e a Tecnologia (FCT, Portugal) through the projects CERN/FIS-NUC/0010/2015 and CFTP-FCT Unit 777 (UID/FIS/00777/2013) which are partially funded through POCTI (FEDER), COMPETE, QREN and EU. GCB was supported in part by the National Science Centre, Poland, the HARMONIA project under contract UMO-2015/18/M/ST00518 (2016-2019). MN acknowledges support from FCT through postdoctoral grant SFRH/BPD/112999/2015.

\clearpage
\appendix
\section{Details on model identification\label{SEC:APP:}}

\subsection{Rows and columns\label{sSEC:APP:RowsColumns}}
Recall the arguments in sections \ref{sSEC:LEFT:ell_i} and \ref{sSEC:RIGHT:r_i}.
Consider the \emph{Left condition}
\begin{equation}
\wNQ{q}=(\ell_1\PR{1}+\ell_2\PR{2}+\ell_3\PR{3})\,\wMQ{q}=\begin{pmatrix}\ell_1&0&0\\ 0&\ell_2&0\\ 0&0&\ell_3\end{pmatrix}\,\wMQ{q},
\end{equation}
and \refEQS{eq:MN:matrices:00:1}--\eqref{eq:MN:matrices:00:2} for $\wMQ{q}$ and $\wNQ{q}$
($q=d, u$)
expressed in terms of the Yukawa matrices
$\Gamma_1, \Gamma_2$ and $\Delta_1, \Delta_2$,
respectively for $q=d$ and $q=u$.
If there were non-zero elements
$\left(\Gamma_1\right)_{ia}\neq 0$ and
$\left(\Gamma_2\right)_{ib}\neq 0$
(or $\left(\Delta_1\right)_{ia}\neq 0$ and $\left(\Delta_2\right)_{ib}\neq 0$ )
in the same row $i$ of both Yukawa matrices, it would follow that
\begin{equation}\label{eq:Cond:Values:00}
\left(\wNQ{q}\right)_{ia}=\ell_i\left(\wMQ{q}\right)_{ia}\,\Rightarrow\,\ell_i=\tb\quad
\text{and}\quad
\left(\wNQ{q}\right)_{ib}=\ell_i\left(\wMQ{q}\right)_{ib}\,\Rightarrow\,\ell_i=-\tbinv\,,
\end{equation}
which is not possible. That is, the rows of the $M_d^0$ and $N_d^0$ matrices
($M_u^0$ and $N_u^0$ matrices) come either
from $\Gamma_1$ or from $\Gamma_2$
(from $\Delta_1$ or from $\Delta_2$), never from both.
In other words, each doublet $Q_{Li}^0$ couples to one and only one doublet $\Hd{j}$.\\
\noindent For the \emph{Right condition}
\begin{equation}
\wNQ{q}=\wMQ{q}\,(r_1\PR{1}+r_2\PR{2}+r_3\PR{3})=\wMQ{q}\,\begin{pmatrix}r_1&0&0\\ 0&r_2&0\\ 0&0&r_3\end{pmatrix},
\end{equation}
it follows similarly that each singlet $d_{Ri}^0$, $u_{Rj}^0$, couples to one and only one doublet $\Hd{k}$.
However, contrary to \emph{Left conditions}, this holds separately for the up and down sectors.
Notice, finally, that the only values that the parameters $\ell_{j}$ and $r_{j}$ can take, following \refEQ{eq:Cond:Values:00} are either $\tb$ or $-\tbinv$.

\subsection{Models\label{sSEC:APP:Classes}}
The models discussed in sections \ref{SEC:LEFT} and \ref{SEC:RIGHT} are representative examples within each class. In the following we briefly comment on some other details on these classes of models.\\
Starting with the BGL models of subsection \ref{mod:BGL}, it is to be noticed that in \refEQ{eq:Lmodel:sym:1} one singles out both the third generation and the down quarks. This leads to a model, the ``bottom'' BGL model, where tree level FCNC are absent in the down sector and are controlled by products of CKM elements $\V{ib}\Vc{jb}$ in the up sector. By choosing for example the second generation, the ``strange'' BGL model, one obtains again tree level FCNC in the up sector but controlled by $\V{is}\Vc{js}$ instead. Furthermore, if instead of the down sector one chooses the up sector, that is
\begin{equation}\label{eq:uBGL:1}
\Hd{2}\mapsto e^{i\theta}\Hd{2},\quad Q^0_{L3}\mapsto e^{i\theta}Q^0_{L3},\quad u_{R3}^0\mapsto e^{i2\theta}u_{R3}^0,\quad \theta\neq 0,\pi.
\end{equation}
instead of \refEQ{eq:Lmodel:sym:1}, one obtains the ``top'' BGL model, with no tree level FCNC in the up sector and FCNC controlled by $\V{ti}\Vc{tj}$ in the down sector. Overall, considering the quark sector alone, there are 6 BGL models, one per quark type.

For all the remaining models, shaped by either \emph{Left} or \emph{Right} conditions, the situation is different. Consider for example the generalised BGL model given by \refEQ{eq:Lmodel:sym:2}. The transformation singles out the third generation with $Q^0_{L3}\mapsto -Q^0_{L3}$. The only trace of that election in \refEQ{eq:Lmodel:Nq:fin:2} is the fact that the unitary vector $\unqvec{q}$ is given by the third row of the unitary matrix $\UqX{q}{L}$. However, if we start with $Q^0_{L2}\mapsto -Q^0_{L2}$ instead, the form of $\mND$ and $\mNU$ remains exactly the same, but with a different interpretation of $\unqvec{q}$ (the second row of $\UqX{q}{L}$ in that case). With $\unqvec{q}$ free to vary -- either $q=d$ or $q=u$, the other fixed via CKM, \refEQ{eq:Lmodel:nund:2} --,
it is clear that the generic parametrization in terms of $\unqvec{q}$
covers simultaneously all three initial possibilities $Q^0_{Lj}\mapsto -Q^0_{Lj}$, $j=1,2,3$. This consideration concerning the generalised BGL model is applicable to the remaining cases: the parametrisation of the $\mND$ and $\mNU$ matrices involving unitary vectors $\unqvec{q}$ or $\rnqvec{q}$ encompasses all initial symmetry assignments. It is to be noticed of course, that despite this fact, the models discussed in different classes are distinct: for example the jBGL model in \refEQ{eq:Lmodel:Nq:fin:3} \emph{cannot} be obtained from the gBGL model in \refEQ{eq:Lmodel:Nq:fin:2} with some election of $\unqvec{q}$; they have a different dependence on $\tb$. The same kind of distinction applies to \refEQ{eq:Rmodel:Nq:fin:1} versus \refEQ{eq:Rmodel:Nq:fin:2} and to \refEQ{eq:Rmodel:Nq:fin:3} versus \refEQ{eq:Rmodel:Nq:fin:4}.

\subsection{Identifying $\Phi_1$ and model discrimination\label{sSEC:APP:12}}
In the most general 2HDM there is nothing to disentangle $\Hd{1}$
from $\Hd{2}$.
Indeed, one can mix them through a unitary transformation
without any physical consequence.
The situation changes once one introduces a symmetry through some
specific form.
We start by noticing that the form of the Abelian symmetry chosen in
eq.~\eqref{eq:Charges:00} already singles out $\Hd{1}$;
it is the field which remains invariant under the symmetry.
Given any generic Abelian symmetry,
this choice can always be made by an appropriate basis transformation
in the space of scalar doublets.
Before that choice is made,
the sub-indices $k=1,2$ in $\Hd{k}$ (and, thus, in $\Yd{k}$,
$\Yu{k}$, and the vevs $v_k$)
are just unphysical labels.
One should notice that models are not yet unequivocally defined,
even after the basis choice is made such that the Abelian symmetry
is expressed as $\Hd{1} \mapsto \Hd{1}$.
This is most easily seen in the simple context of the
$\ZZ{2}$ Natural Flavour Conservation models of
Glashow-Weinberg \cite{Glashow:1976nt}.
In that context,
after a scalar basis choice is made such that the scalars transform as
$\Hd{1} \mapsto \Hd{1}$ and $\Hd{2} \mapsto - \Hd{2}$,
one can still choose for the right handed quarks
the transformations
(the same for all quarks of a given charge)
\begin{alignat}{3}
& d_R \mapsto d_R\, ,\qquad
&&
u_R \mapsto u_R\, ;
\\
& d_R \mapsto - d_R\, ,\qquad
&&
u_R \mapsto - u_R\, ;
\\
& d_R \mapsto d_R\, ,\qquad
&&
u_R \mapsto - u_R\, ;
\\
& d_R \mapsto - d_R\, ,\qquad
&&
u_R \mapsto u_R\, .
\end{alignat}
In the first two equations, the up and down quarks couple to the same
field (be it $\Hd{1}$ or $\Hd{2}$; it does not matter).
This is known as Type I.
In the last two equations, the up and down quarks couple to the different
fields; which is known as Type II.
Denoting a field by $\Hd{1}$ or $\Hd{2}$ has no physical meaning.
The most direct counting can be obtained by choosing (say) $\Hd{2}$
as the field which couples to the up quarks.
This is what attributes physical meaning to the labels $1$ and $2$.
With this choice,
the sub-indices of the Yukawa matrices ($\Yd{k}$ and
$\Yu{k}$) acquire physical meaning.
The same happens with the vevs $v_k$ \cite{Haber:2006ue}.
Subsequent changes in the basis for fermions will alter the form
of the Yukawa matrices, but not their rank.

A similar analysis can be made for the models discussed in this paper,
except that here the right handed up quarks do not couple all
to the same doublet.
However,
as can be seen from the form of the matrices shown,
$\textrm{rank}\left(\Gamma_1\right) + \textrm{rank}\left(\Gamma_2\right) = 3$
and
$\textrm{rank}\left(\Delta_1\right) + \textrm{rank}\left(\Delta_2\right) = 3$.
As a result,
one can define physically the label in $\Phi_1$
as the scalar which couples to most of the up quarks.
All subsequent choices are physically meaningful.
Alternatively,
one can define $\Phi_1$ as the field which obeys
$\Phi_1 \rightarrow \Phi_1$ under the Abelian symmetry,
at the price of an apparent but illusory doubling of the number of model types.
This is shown explicitly for the \emph{Right} models in
Table~\ref{TAB:countRightModels}
\footnote{Due to eq.~\eqref{eq:Cond:Values:00}, the change
$1 \leftrightarrow 2$ implies a change
$t_\beta \leftrightarrow - t_\beta^{-1}$ in the parametrization
of the $N_q$ matrices. Thus, models which could seem to differ
by such a change, do in fact correspond to the same model.}.
\begin{table}[h!tb]
\begin{center}
{\renewcommand{\arraystretch}{1.2}%
\begin{tabular}{|c|c|c|c|c|}
\hline
\backslashbox{{\small rank $\Gamma_k$}}{{\small rank $\Delta_k$}} & 3,0 & 0,3 & 2,1 & 1,2
\\ \hline
3,0 &  Type I & Type II & Type A & Type B
\\ \hline
0,3 &  Type II & Type I & Type B & Type A
\\ \hline
2,1 & Type C & Type D & Type E & Type F
\\ \hline
1,2 & Type D & Type C & Type F & Type E
\\ \hline
\end{tabular}
}
\caption{Identification of the \emph{Right} models (and the usual
Type I and Type II), in terms of the ranks of the Yukawa matrices,
in the order $\Delta_1, \Delta_2$ (in columns),
and $\Gamma_1, \Gamma_2$ (in rows).\label{TAB:countRightModels}}
\end{center}
\end{table}

In this analysis, we have used the fact that, if for example
$\Delta_1$ has two columns and $\Delta_2$ the third,
it is immaterial their placement and, moreover,
their placement with respect to the placement of the columns
which appear in $\Gamma_1$ and $\Gamma_2$.
To be specific, let us consider the Type A matrices of
eq.~\eqref{eq:Rmodel:Y:1},
where we have chosen $\Delta_1$ to have the first two columns
and $\Delta_2$ the last, while $\Gamma_1$ has all columns.
We could have chosen $\Delta_1$ to have the first and last column,
with $\Delta_2$ having the second column.
The different permutations refer only to the labels in the
space of right handed up quarks (which is completely
detached from the space of right handed down quarks).
Such choices are indistinguishable.

The situation is easier for \emph{Left} models,
because left up quark and left down quark fields belong
to the same doublet, leading to the restriction in
eq.~\eqref{eq:Lmodel:nund:2}.
Hence,
as seen in section~\ref{SEC:LEFT},
the possible ranks of $(\Gamma_1 , \Gamma_2)$ are only
$(1,2)$ and $(2,1)$.
Thus,
instead of
Table~\ref{TAB:countRightModels}
one obtains the much simpler
Table~\ref{TAB:countLeftModels}.
\begin{table}[h!tb]
\begin{center}
{\renewcommand{\arraystretch}{1.2}%
\begin{tabular}{|c|c|c|}
\hline
\backslashbox{{\small rank $\Gamma_k$}}{{\small rank $\Delta_k$}} & 2,1 & 1,2
\\ \hline
2,1 & gBGL & jBGL
\\ \hline
1,2 & jBGL & gBGL
\\ \hline
\end{tabular}
}
\caption{Identification of the \emph{Left} models
in terms of the ranks of the Yukawa matrices,
in the order $\Delta_1, \Delta_2$ (in columns),
and $\Gamma_1, \Gamma_2$ (in rows).\label{TAB:countLeftModels}}
\end{center}
\end{table}
We are now ready to develop basis invariant conditions for the determination
of the various models.

\subsection{Invariant conditions\label{sSEC:APP:Conditions}}
Here,
we present conditions for the identification of the various types of models
discussed in this article,
which are invariant under basis transformations in
the spaces of left-handed doublets and of up-type and down-type
right-handed singlets.
For BGL and generalised BGL models, the following matrix conditions hold
\cite{Alves:2017xmk}:
\begin{alignat}{2}
&\text{BGL models: }&\Ydd{1}\Yd{2}=0,\quad \Yud{1}\Yu{2}=0,\quad \Ydd{1}\Yu{2}=0,\quad \Ydd{2}\Yu{1}=0,\nonumber\\
& &\text{and}\ \ \Yd{1}\Ydd{2}=0\ \text{(dBGL)}\ \ \text{or}
\ \ \Yu{1}\Yud{2}=0\ \text{(uBGL)},\nonumber\\*[2mm]
&\text{gBGL models: }&\Ydd{1}\Yd{2}=0,\quad \Yud{1}\Yu{2}=0,\quad \Ydd{1}\Yu{2}=0,\quad \Ydd{2}\Yu{1}=0.
\label{eq:BGLbBGL:Cond:00}
\end{alignat}
Their importance resides in the fact that, under a weak basis transformation (WBT) of the fermion fields
\begin{equation}\label{eq:WBT:fermions:00}
Q_L\mapsto \WL\, Q_L,\quad d_R\mapsto \WdR\, d_R,\quad u_R\mapsto \WuR\,u_R,\quad \WL,\WdR,\WuR\in U(3),
\end{equation}
the Yukawa coupling matrices are transformed as
\begin{equation}\label{eq:WBT:Yukawas:00}
\Yd{i}\mapsto \WLd\,\Yd{i}\,\WdR,\quad \Yu{i}\mapsto \WLd\,\Yu{i}\,\WuR\,,
\end{equation}
and, although the WBT in \refEQS{eq:WBT:fermions:00}--\eqref{eq:WBT:Yukawas:00} may hide the symmetry under the Abelian transformations in \refEQ{eq:Charges:00}, the conditions in \refEQS{eq:BGLbBGL:Cond:00} are in any case invariant. In general, the different combinations of $\Yd{i}$, $\Yu{j}$, which are invariant under some of the WBT are the following.
\begin{itemize}
\item Invariant under $\WL$ WBT,
\begin{align}
\Ydd{i}\Yd{j}&\mapsto \WdRd\,\Ydd{i}\Yd{j}\,\WdR,\nonumber\\
\Yud{i}\Yu{j}&\mapsto \WuRd\,\Yud{i}\Yu{j}\,\WuR,\nonumber\\
\Ydd{i}\Yu{j}&\mapsto \WdRd\,\Ydd{i}\Yu{j}\,\WuR,
\label{eq:WBT:WL:inv:00}
\end{align}
(and, of course, $\Yud{i}\Yd{j}=(\Ydd{j}\Yu{i})^\dagger$)
\item Invariant under $\WdR$ and $\WuR$ WBT,
\begin{align}
\Yd{i}\Ydd{j}&\mapsto \WLd\,\Yd{i}\Ydd{j}\,\WL,\nonumber\\
\Yu{i}\Yud{j}&\mapsto \WLd\,\Yu{i}\Yud{j}\,\WL.
\label{eq:WBT:WqR:inv:00}
\end{align}
\end{itemize}
Considering in addition the \emph{Left} and \emph{Right conditions}
of eqs.~\eqref{eq:LeftProjection:00} and \eqref{eq:RightProjection:00}, respectively,
we can straightforwardly obtain invariant conditions.
This is what we turn to next.
\subsubsection{Left conditions\label{ssSEC:APP:LeftCond}}
In terms of the Yukawa matrices, the \emph{Left conditions} are
\begin{align}\label{eq:Yukawas:LCond:00}
\Gamma_1
&=
e^{- i\xi_1}\frac{\sqrt{2}}{v}\sb(\tbinv\id+\wLP{d})\wMQ{d},
&
\Gamma_2
&=
- e^{- i\xi_2}\frac{\sqrt{2}}{v}\cb(-\tb\id+\wLP{d})\wMQ{d},
\\
\Delta_1
&=
e^{i\xi_1}\frac{\sqrt{2}}{v}\sb(\tbinv\id+\wLP{u})\wMQ{u},
&
\Delta_2
&=
- e^{i\xi_2}\frac{\sqrt{2}}{v}\cb(-\tb\id+\wLP{u})\wMQ{u},
\end{align}
where we have used eqs.~\eqref{eq:MN:matrices:00:1} and \eqref{eq:MN:matrices:00:2}.
Then,
\begin{equation}
\Ydd{1}\Yd{2}=-\sb\cb e^{i\xi}\frac{2}{v^2}\wMDd(\tbinv\id+\wLP{d})(-\tb\id+\wLP{d})\wMD\,,
\end{equation}
where
\begin{equation}
(\tbinv\id+\wLP{d})(-\tb\id+\wLP{d})=\sum_{j=1}^3(\ell_j^{[d]}+\tbinv)(\ell_j^{[d]}-\tb)\PR{j}=0
\end{equation}
since $\ell_j^{[d]}$ is equal either to $-\tbinv$ or to $\tb$, thus giving $\Ydd{1}\Yd{2}=0$. For the up Yukawa matrices, the conclusion is identical, and thus the conditions in \refEQS{eq:WBT:WL:inv:00} involving separately the up and down quark sectors are trivially
\begin{equation}\label{eq:Left:Cond:00}
\Ydd{1}\Yd{2}=0,\quad \Yud{1}\Yu{2}=0\,.
\end{equation}
For the conditions involving Yukawa matrices from both sectors,
proceeding along similar lines, one finds
\begin{align}
e^{-i2\xi_1}\frac{v^2}{2}\Ydd{1}\Yu{1}&=\sb^2\,\wMDd\left[\sum_{j=1}^3(\ell_j^{[d]}+\tbinv)(\ell_j^{[u]}+\tbinv)\PR{j}\right]\wMU,\nonumber\\
-e^{-i(\xi_1+\xi_2)}\frac{v^2}{2}\Ydd{1}\Yu{2}&=\sb\cb\,\wMDd\left[\sum_{j=1}^3(\ell_j^{[d]}+\tbinv)(\ell_j^{[u]}-\tb)\PR{j}\right]\wMU,\nonumber\\
-e^{-i(\xi_1+\xi_2)}\frac{v^2}{2}\Ydd{2}\Yu{1}&=\sb\cb\,\wMDd\left[\sum_{j=1}^3(\ell_j^{[d]}-\tb)(\ell_j^{[u]}+\tbinv)\PR{j}\right]\wMU,\nonumber\\
e^{-i2\xi_2}\frac{v^2}{2}\Ydd{2}\Yu{2}&=\cb^2\,\wMDd\left[\sum_{j=1}^3(\ell_j^{[d]}-\tb)(\ell_j^{[u]}-\tb)\PR{j}\right]\wMU,
\label{eq:Left:Cond:01}
\end{align}
and one can readily obtain the additional conditions
\begin{alignat}{2}
&\text{BGL and gBGL models: }\ \ &\Ydd{1}\Yu{2}=0,\quad \Ydd{2}\Yu{1}=0,
\nonumber\\*[2mm]
&\text{jBGL models: }&\Ydd{1}\Yu{1}=0,\quad \Ydd{2}\Yu{2}=0.
\label{eq:Left:Cond:02}
\end{alignat}
The remaining matrix products, including $\Yd{1}\Ydd{2}$, $\Yu{1}\Yud{2}$, are different from $0$ and do not give invariant conditions like \refEQS{eq:Left:Cond:00} and \eqref{eq:Left:Cond:02}.
\subsubsection{Right conditions\label{ssSEC:APP:RightCond}}
For models with \emph{Right conditions}, the analog of \refEQ{eq:Left:Cond:00} is simply
\begin{equation}\label{eq:Right:Cond:00}
\Yd{1}\Ydd{2}=0,\quad \Yu{1}\Yud{2}=0\,.
\end{equation}
One could naively think that conditions such as $\Gamma_2 \Delta_1^\dagger$
could be used to distinguish among different models.
But, such conditions cannot be used, for they are not covariant under WBT.
Fortunately,
the different \emph{Right} models can be distinguished in a basis invariant way
by the rank of the $\Gamma_1$ and $\Delta_1$ matrices.
\subsubsection{Summary\label{ssSEC:APP:AllCond}}
We summarize in Table~\ref{TAB:InvCondAllModels} the invariant conditions
associated with all models discussed in this article.
\begin{table}[h!tb]
\begin{center}
{\renewcommand{\arraystretch}{1.4}%
\begin{tabular}{|c|c|}
\hline
Model & Invariant Conditions \\ \hline\hline
{dBGL}
& $\Ydd{1}\Yd{2}=0$, $\Yud{1}\Yu{2}=0$, $\Ydd{1}\Yu{2}=0$, $\Ydd{2}\Yu{1}=0$, $\Yd{1}\Ydd{2}=0$\\ \hline
{uBGL}
& $\Ydd{1}\Yd{2}=0$, $\Yud{1}\Yu{2}=0$, $\Ydd{1}\Yu{2}=0$, $\Ydd{2}\Yu{1}=0$, $\Yu{1}\Yud{2}=0$\\ \hline
gBGL & $\Ydd{1}\Yd{2}=0$, $\Yud{1}\Yu{2}=0$, $\Ydd{1}\Yu{2}=0$, $\Ydd{2}\Yu{1}=0$\\ \hline
jBGL & $\Ydd{1}\Yd{2}=0$, $\Yud{1}\Yu{2}=0$, $\Ydd{1}\Yu{1}=0$, $\Ydd{2}\Yu{2}=0$\\ \hline\hline
{Type A}
& $\Yd{1}\Ydd{2}=0$, $\Yu{1}\Yud{2}=0$, $\text{rank}(\Yd{1})=3$, $\text{rank}(\Yu{1})=2$\\ \hline
{Type B}
& $\Yd{1}\Ydd{2}=0$, $\Yu{1}\Yud{2}=0$, $\text{rank}(\Yd{1})=3$, $\text{rank}(\Yu{1})=1$\\ \hline
{Type C}
& $\Yd{1}\Ydd{2}=0$, $\Yu{1}\Yud{2}=0$, $\text{rank}(\Yd{1})=2$, $\text{rank}(\Yu{1})=3$\\ \hline
{Type D}
& $\Yd{1}\Ydd{2}=0$, $\Yu{1}\Yud{2}=0$, $\text{rank}(\Yd{1})=2$, $\text{rank}(\Yu{1})=0$\\ \hline
{Type E}
& $\Yd{1}\Ydd{2}=0$, $\Yu{1}\Yud{2}=0$, $\text{rank}(\Yd{1})=2$, $\text{rank}(\Yu{1})=2$\\ \hline
{Type F}
& $\Yd{1}\Ydd{2}=0$, $\Yu{1}\Yud{2}=0$, $\text{rank}(\Yd{1})=2$, $\text{rank}(\Yu{1})=1$\\ \hline
\end{tabular}
}
\caption{Summary of invariant conditions.\label{TAB:InvCondAllModels}}
\end{center}
\end{table}

\clearpage

\begin{thebibliography}{10}

\bibitem{Lee:1973iz}
T.~Lee, {\it {A Theory of Spontaneous T Violation}},  {\em Phys.Rev.} {\bf D8}
  (1973) 1226--1239.

\bibitem{Branco:2011iw}
G.~Branco, P.~Ferreira, L.~Lavoura, M.~Rebelo, M.~Sher, {\em et~al.}, {\it
  {Theory and phenomenology of two-Higgs-doublet models}},  {\em Phys.Rept.}
  {\bf 516} (2012) 1--102, [\href{http://xxx.lanl.gov/abs/1106.0034}{{\tt
  1106.0034}}].

\bibitem{Ivanov:2017dad}
I.~P. Ivanov, {\it {Building and testing models with extended Higgs sectors}},
  {\em Prog. Part. Nucl. Phys.} {\bf 95} (2017) 160--208,
  [\href{http://xxx.lanl.gov/abs/1702.03776}{{\tt 1702.03776}}].

\bibitem{Glashow:1976nt}
S.~L. Glashow and S.~Weinberg, {\it {Natural Conservation Laws for Neutral
  Currents}},  {\em Phys.Rev.} {\bf D15} (1977) 1958.

\bibitem{Weinberg:1976hu}
S.~Weinberg, {\it {Gauge Theory of CP Violation}},  {\em Phys. Rev. Lett.} {\bf
  37} (1976) 657.

\bibitem{Branco:1979pv}
G.~C. Branco, {\it {Spontaneous CP Violation in Theories with More Than Four
  Quarks}},  {\em Phys. Rev. Lett.} {\bf 44} (1980) 504.

\bibitem{Branco:2015bfb}
G.~C. Branco and I.~P. Ivanov, {\it {Group-theoretic restrictions on generation
  of CP-violation in multi-Higgs-doublet models}},  {\em JHEP} {\bf 01} (2016)
  116, [\href{http://xxx.lanl.gov/abs/1511.02764}{{\tt 1511.02764}}].

\bibitem{Branco:1996bq}
G.~Branco, W.~Grimus, and L.~Lavoura, {\it {Relating the scalar flavor changing
  neutral couplings to the CKM matrix}},  {\em Phys.Lett.} {\bf B380} (1996)
  119--126, [\href{http://xxx.lanl.gov/abs/hep-ph/9601383}{{\tt
  hep-ph/9601383}}].

\bibitem{Botella:2009pq}
F.~Botella, G.~Branco, and M.~Rebelo, {\it {Minimal Flavour Violation and
  Multi-Higgs Models}},  {\em Phys.Lett.} {\bf B687} (2010) 194--200,
  [\href{http://xxx.lanl.gov/abs/0911.1753}{{\tt 0911.1753}}].

\bibitem{Botella:2011ne}
F.~Botella, G.~Branco, M.~Nebot, and M.~Rebelo, {\it {Two-Higgs Leptonic
  Minimal Flavour Violation}},  {\em JHEP} {\bf 1110} (2011) 037,
  [\href{http://xxx.lanl.gov/abs/1102.0520}{{\tt 1102.0520}}].

\bibitem{Botella:2014ska}
F.~Botella, G.~Branco, A.~Carmona, M.~Nebot, L.~Pedro, and M.~Rebelo, {\it
  {Physical Constraints on a Class of Two-Higgs Doublet Models with FCNC at
  tree level}},  {\em JHEP} {\bf 1407} (2014) 078,
  [\href{http://xxx.lanl.gov/abs/1401.6147}{{\tt 1401.6147}}].

\bibitem{Bhattacharyya:2014nja}
G.~Bhattacharyya, D.~Das, and A.~Kundu, {\it {Feasibility of light scalars in a
  class of two-Higgs-doublet models and their decay signatures}},  {\em
  Phys.Rev.} {\bf D89} (2014) 095029,
  [\href{http://xxx.lanl.gov/abs/1402.0364}{{\tt 1402.0364}}].

\bibitem{Botella:2015hoa}
F.~J. Botella, G.~C. Branco, M.~Nebot, and M.~N. Rebelo, {\it {Flavour Changing
  Higgs Couplings in a Class of Two Higgs Doublet Models}},  {\em Eur. Phys.
  J.} {\bf C76} (2016), no.~3 161,
  [\href{http://xxx.lanl.gov/abs/1508.05101}{{\tt 1508.05101}}].

\bibitem{Alves:2017xmk}
J.~M. Alves, F.~J. Botella, G.~C. Branco, F.~Cornet-Gomez, and M.~Nebot, {\it
  {Controlled Flavour Changing Neutral Couplings in Two Higgs Doublet Models}},
   {\em Eur. Phys. J.} {\bf C77} (2017), no.~9 585,
  [\href{http://xxx.lanl.gov/abs/1703.03796}{{\tt 1703.03796}}].

\bibitem{Ferreira:2010ir}
P.~M. Ferreira and J.~P. Silva, {\it {Abelian symmetries in the
  two-Higgs-doublet model with fermions}},  {\em Phys. Rev.} {\bf D83} (2011)
  065026, [\href{http://xxx.lanl.gov/abs/1012.2874}{{\tt 1012.2874}}].

\bibitem{Serodio:2013gka}
H.~Serôdio, {\it {Yukawa sector of Multi Higgs Doublet Models in the presence
  of Abelian symmetries}},  {\em Phys. Rev.} {\bf D88} (2013), no.~5 056015,
  [\href{http://xxx.lanl.gov/abs/1307.4773}{{\tt 1307.4773}}].

\bibitem{Georgi:1978ri}
H.~Georgi and D.~V. Nanopoulos, {\it {Suppression of Flavor Changing Effects
  From Neutral Spinless Meson Exchange in Gauge Theories}},  {\em Phys. Lett.}
  {\bf B82} (1979) 95.

\bibitem{Donoghue:1978cj}
J.~F. Donoghue and L.~F. Li, {\it {Properties of Charged Higgs Bosons}},  {\em
  Phys. Rev.} {\bf D19} (1979) 945.

\bibitem{Botella:1994cs}
F.~J. Botella and J.~P. Silva, {\it {Jarlskog - like invariants for theories
  with scalars and fermions}},  {\em Phys. Rev.} {\bf D51} (1995) 3870--3875,
  [\href{http://xxx.lanl.gov/abs/hep-ph/9411288}{{\tt hep-ph/9411288}}].

\bibitem{Haber:1978jt}
H.~E. Haber, G.~L. Kane, and T.~Sterling, {\it {The Fermion Mass Scale and
  Possible Effects of Higgs Bosons on Experimental Observables}},  {\em Nucl.
  Phys.} {\bf B161} (1979) 493--532.

\bibitem{Hall:1981bc}
L.~J. Hall and M.~B. Wise, {\it {Flavor changing Higgs - boson couplings}},
  {\em Nucl. Phys.} {\bf B187} (1981) 397.

\bibitem{Barger:1989fj}
V.~D. Barger, J.~L. Hewett, and R.~J.~N. Phillips, {\it {New Constraints on the
  Charged Higgs Sector in Two Higgs Doublet Models}},  {\em Phys. Rev.} {\bf
  D41} (1990) 3421--3441.

\bibitem{Joshipura:1990pi}
A.~S. Joshipura and S.~D. Rindani, {\it {Naturally suppressed flavor violations
  in two Higgs doublet models}},  {\em Phys.Lett.} {\bf B260} (1991) 149--153.

\bibitem{Joshipura:1990xm}
A.~S. Joshipura, {\it {Neutral Higgs and CP violation}},  {\em Mod. Phys.
  Lett.} {\bf A6} (1991) 1693--1700.

\bibitem{Lavoura:1994ty}
L.~Lavoura, {\it {Models of CP violation exclusively via neutral scalar
  exchange}},  {\em Int.J.Mod.Phys.} {\bf A9} (1994) 1873--1888.

\bibitem{Branco:1985aq}
G.~C. Branco and M.~N. Rebelo, {\it {The Higgs Mass in a Model With Two Scalar
  Doublets and Spontaneous {CP} Violation}},  {\em Phys. Lett.} {\bf B160}
  (1985) 117--120.

\bibitem{Fontes:2014xva}
D.~Fontes, J.~C. Romão, and J.~P. Silva, {\it {$h \rightarrow Z \gamma$ in the
  complex two Higgs doublet model}},  {\em JHEP} {\bf 12} (2014) 043,
  [\href{http://xxx.lanl.gov/abs/1408.2534}{{\tt 1408.2534}}].

\bibitem{Grzadkowski:2016szj}
B.~Grzadkowski, O.~M. Ogreid, and P.~Osland, {\it {Spontaneous CP violation in
  the 2HDM: physical conditions and the alignment limit}},  {\em Phys. Rev.}
  {\bf D94} (2016), no.~11 115002,
  [\href{http://xxx.lanl.gov/abs/1609.04764}{{\tt 1609.04764}}].

\bibitem{Fontes:2017zfn}
D.~Fontes, M.~Mühlleitner, J.~C. Romão, R.~Santos, J.~P. Silva, and
  J.~Wittbrodt, {\it {The C2HDM revisited}},  {\em JHEP} {\bf 02} (2018) 073,
  [\href{http://xxx.lanl.gov/abs/1711.09419}{{\tt 1711.09419}}].

\bibitem{Khachatryan:2016vau}
{\bf ATLAS, CMS} Collaboration, G.~Aad {\em et~al.}, {\it {Measurements of the
  Higgs boson production and decay rates and constraints on its couplings from
  a combined ATLAS and CMS analysis of the LHC pp collision data at $
  \sqrt{s}=7 $ and 8 TeV}},  {\em JHEP} {\bf 08} (2016) 045,
  [\href{http://xxx.lanl.gov/abs/1606.02266}{{\tt 1606.02266}}].

\bibitem{Blankenburg:2012ex}
G.~Blankenburg, J.~Ellis, and G.~Isidori, {\it {Flavour-Changing Decays of a
  125 GeV Higgs-like Particle}},  {\em Phys. Lett.} {\bf B712} (2012) 386--390,
  [\href{http://xxx.lanl.gov/abs/1202.5704}{{\tt 1202.5704}}].

\bibitem{Aad:2014dya}
{\bf ATLAS} Collaboration, G.~Aad {\em et~al.}, {\it {Search for top quark
  decays $t \to qH$ with $H \to \gamma\gamma$ using the ATLAS detector}},  {\em
  JHEP} {\bf 06} (2014) 008, [\href{http://xxx.lanl.gov/abs/1403.6293}{{\tt
  1403.6293}}].

\bibitem{Aaboud:2017mfd}
{\bf ATLAS} Collaboration, M.~Aaboud {\em et~al.}, {\it {Search for top quark
  decays $t\rightarrow qH$, with $H\to\gamma\gamma$, in $\sqrt{s}=13$ TeV $pp$
  collisions using the ATLAS detector}},  {\em JHEP} {\bf 10} (2017) 129,
  [\href{http://xxx.lanl.gov/abs/1707.01404}{{\tt 1707.01404}}].

\bibitem{Khachatryan:2014jya}
{\bf CMS} Collaboration, V.~Khachatryan {\em et~al.}, {\it {Searches for heavy
  Higgs bosons in two-Higgs-doublet models and for $t\to ch$ decay using
  multilepton and diphoton final states in $pp$ collisions at 8 TeV}},  {\em
  Phys. Rev.} {\bf D90} (2014) 112013,
  [\href{http://xxx.lanl.gov/abs/1410.2751}{{\tt 1410.2751}}].

\bibitem{Khachatryan:2016atv}
{\bf CMS} Collaboration, V.~Khachatryan {\em et~al.}, {\it {Search for top
  quark decays via Higgs-boson-mediated flavor-changing neutral currents in pp
  collisions at $ \sqrt{s}=8 $ TeV}},  {\em JHEP} {\bf 02} (2017) 079,
  [\href{http://xxx.lanl.gov/abs/1610.04857}{{\tt 1610.04857}}].

\bibitem{Nebot:2015wsa}
M.~Nebot and J.~P. Silva, {\it {Self-cancellation of a scalar in neutral meson
  mixing and implications for the LHC}},  {\em Phys. Rev.} {\bf D92} (2015),
  no.~8 085010, [\href{http://xxx.lanl.gov/abs/1507.07941}{{\tt 1507.07941}}].

\bibitem{Haber:2006ue}
H.~E. Haber and D.~O'Neil, {\it {Basis-independent methods for the
  two-Higgs-doublet model. II. The Significance of tan$\beta$}},  {\em Phys.
  Rev.} {\bf D74} (2006) 015018,
  [\href{http://xxx.lanl.gov/abs/hep-ph/0602242}{{\tt hep-ph/0602242}}].
  [Erratum: Phys. Rev.D74,no.5,059905(2006)].

\end{thebibliography}
\providecommand{\href}[2]{#2}\begingroup\raggedright\endgroup

\end{document}